\documentclass{nature_dani_sp}
\usepackage{amssymb,amsfonts,amsmath,rotating}
\usepackage{lscape}
\usepackage{hyperref}
\usepackage{graphicx}
\usepackage[font=small]{caption}

\usepackage{color}

\title{Learning-Induced Autonomy of Sensorimotor Systems}

\author{Danielle S. Bassett$^{1,2,*}$, Muzhi Yang$^{1,3}$, Nicholas F.
Wymbs$^{4}$, Scott T. Grafton$^{4}$}

\begin{document}

\maketitle

\begin{affiliations}
 \item Complex Systems Group, Department of Bioengineering, University of Pennsylvania, Philadelphia, PA, 19104, USA
 \item Department of Electrical Engineering, University of Pennsylvania, Philadelphia, PA, 19104, USA
 \item Applied Mathematics and Computational Science Graduate Group, University of Pennsylvania, Philadelphia, PA, 19104, USA
 \item Department of Psychological and Brain Sciences and UCSB Brain Imaging Center, University of California, Santa Barbara, CA 93106, USA
\end{affiliations}


\newpage
\begin{abstract}
 Distributed networks of brain areas interact with one another in a time-varying fashion to enable complex cognitive and sensorimotor functions. Here we use novel network analysis algorithms to test the recruitment and integration of large-scale functional neural circuitry during learning. Using functional magnetic resonance imaging (fMRI) data acquired from healthy human participants, from initial training through mastery of a simple motor skill, we investigate changes in the architecture of functional connectivity patterns that promote learning. Our results reveal that learning induces an autonomy of sensorimotor systems and that the release of cognitive control hubs in frontal and cingulate cortices predicts individual differences in the rate of learning on other days of practice. Our general statistical approach is applicable across other cognitive domains and provides a key to understanding time-resolved interactions between distributed neural circuits that enable task performance.
\end{abstract}

\newpage

Human skill learning is a complex phenomenon requiring flexibility to adapt existing brain function to drive desired behavior \cite{Ajemian2013}. Such adaptibility requires a fine-scale control of distributed neural circuits to transform a skill from a slow and challenging activity to one that is fast and automatic \cite{Grafton2007}. However, the dynamic manipulation and control of these circuits over the course of learning remains sparsely studied \cite{Rowe2012}.

Previous learning experiments over short time intervals have clearly demonstrated that a few pairwise functional interactions between motor cortical areas -- such as primary motor cortex and supplementary motor cortex -- strengthen with skill acquisition \cite{Sun2007,Xiong2009,Buchel1999}. However, these local interactions sit within the larger context of the whole brain, which is teeming with thousands or millions of changing functional interactions between hundreds or thousands of large cortical structures. The reconfiguration of this wider dynamic network must be involved in shaping learning over the course of training.

Methods for studying the dynamics of distributed and integrated circuits during cognitive processes remain limited and underpowered \cite{Fedorenko2014}. Prior work has relied on studies of specific brain regions or neurons, and on studies of static network representations of structural or functional circuits or connectomes \cite{Bassett2006b,Bassett2009b,Bassett2010b,Bullmore2011,Sporns2010}, both of which are unable to describe temporal dependencies between spatially segregated neural circuits. Initial efforts to utilize both network representations of circuitry \emph{and} analyses sensitive to dynamics have identified gross summary statistics of the whole brain (e.g., network flexibility) \cite{Bassett2011b,Bassett2013a}, which preclude a specific characterization of individual circuits and their evolution.

Indeed, to date there are no statistically robust methods to reliably identify functional modules over short time intervals and characterize their changes over time in individual subjects. We introduce a method for the first time that allows us to address this gap. Here, we use \emph{dynamic network neuroscience} to probe learning-related changes in inter-regional communication patterns, and to link these changes to individual differences in behavioral outcomes. Building on the formalism of network science \cite{Holme2011}, this approach harnesses a new set of tools from applied mathematics \cite{Kivela2013} to treat the patterns of communication between brain regions as evolving networks \cite{Bassett2011b,Bassett2012b,Bassett2013a,Bassett2013b,Doron2012}, whose reconfiguration dynamics are tightly tied to observable changes in behavior.

We use a dynamic network analysis to consider three basic questions about how whole brain functional networks reconfigure during skill learning \cite{Bassett2011b,Mantzaris2013,Bassett2013a,Bassett2013c,Bassett2012b}. First, are there sets of brain regions (or \emph{modules}) that preferentially interact with one another during task performance, and if so, do these modules or their interactions change with learning? Basic intuition suggests that motor output areas might form a task-relevant module. Because sensory guided behavior becomes less of a critical driver for some motor skills in late learning \cite{Logan1988}, we might also hypothesize that sensory regions form a task-relevant module, whose integration with the motor modules changes with training. Second, what role does association cortex play in task performance? During early learning, performance can be enhanced by explicit strategies, cognitive control, and guided attention, but with extensive practice these enhancements are no longer needed \cite{Hikosaka2002,Petersen1998}. One might therefore expect that the interaction between associative and motor modules changes with training. Third, could the interplay between task-relevant (e.g., motor and sensory) modules or the involvement of association cortex explain the dramatic differences between individual people in their capacity to learn? Given that initial skill acquisition requires heightened attentional demands \cite{Hikosaka2002,Petersen1998}, one might hypothesize that slow learners rely on associative systems more or for longer periods of time than fast learners.

We address these questions by quantifying changes in putative functional modules induced by motor learning. We build temporal networks from fMRI data acquired alongside motor sequence learning by (i) subdividing the brain into 112 cortical and subcortical areas (nodes), and (ii) calculating the functional connectivity (edges) between pairs of regions in independent time windows (2--3 minutes in duration corresponding to trial blocks) during task performance (Fig.~\ref{fig:methods}A). Using a clustering approach designed for time-evolving networks \cite{Mucha2010}, we extract groups of brain regions (network communities) that are coherently active in each time window, with each group putatively responsible for a unique cognitive function \cite{Power2011} (Fig.~\ref{fig:methods}B). Using this approach, we obtain subject-specific communities at multiple trial blocks over the course of 6 weeks (4 scan sessions) of sequence learning, during which subjects practiced a set of sequences at 3 levels of training intensity (extensive, EXT; moderate, MOD; minimal, MIN) (Fig.~\ref{fig:methods}C). This construction allows us to ask how large-scale systems adapt during motor skill acquisition, and to identify characteristics of that adaptation that predict individual differences in learning.

\section*{Results}
\subsection{Summary Architecture of Learning}
We first sought to address the question: ``Are there sets of brain regions (or \emph{modules}) that preferentially interact with one another during task performance?'' To answer this question, we examined the brain network architecture that was consistently expressed across the motor skill learning experiment \footnote{We note that this task-based architecture likely differs significantly from task-free architecture.}. For each pair of brain regions, we computed a normalized value of \emph{module allegiance}, which represents the probability that area $i$ and area $j$ have been assigned to the same functional community by time-resolved clustering methods \cite{Mucha2010,Bassett2011b} (see Methods). More specifically, we construct a matrix $\mathbf{T}$ whose elements $T_{ij}$ indicate the number of times that nodes $i$ and $j$ have been assigned to the same community, in the set of functional brain networks constructed from all subjects, scanning sessions, sequence types, and trial blocks. We then compute $\mathbf{P} = \frac{1}{C}\mathbf{T}$, where $C$ is total number of partitions, to obtain the module allegiance matrix $\mathbf{P}$ whose elements $P_{ij}$ give the probability that area $i$ and $j$ are in the same community (Fig.~\ref{fig:methods}D).

\begin{figure}[h]
\centerline{\includegraphics[width=0.65\textwidth]{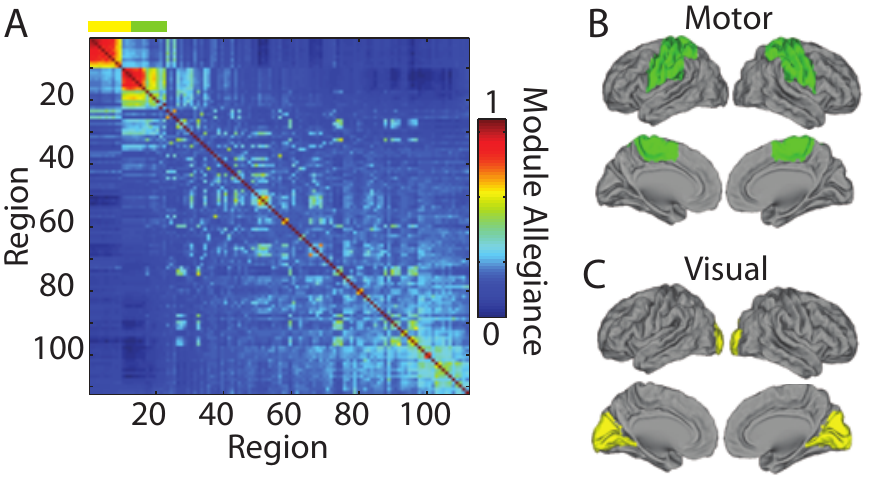}}
\caption{\textbf{Summary Architecture of Learning} \emph{(A)} The module allegiance matrix that indicates for each pair of nodes the probability that those two nodes are located in the same functional community over subjects, scanning sessions, sequence types, and trial blocks. This module allegiance matrix displays two putative functional modules composed of brain regions that are consistently grouped into the same network community: \emph{(B)} one composed of primary and secondary sensorimotor areas, and \emph{(C)} one composed of primary visual cortex. Brain regions in panel \emph{(A)} have been ordered to maximize strong associations along the diagonal.} \label{fig:stable}
\end{figure}

The module allegiance matrix is a summary of the brain network architecture accompanying learning. This architecture displays several interesting anatomical features (Fig.~\ref{fig:stable}A), including two sets of brain regions that are consistently grouped into the same network community. We refer to these two sets as \emph{putative functional modules}. One of these modules is composed of primary and secondary sensorimotor regions (Fig.~\ref{fig:stable}B), and the other module is composed of primary visual cortex (Fig.~\ref{fig:stable}C). See Table~\ref{Tab1} for region labels associated with the two modules. The dissociation of these two modules indicates that brain areas in these two systems display significantly different time courses of BOLD activation.

We will refer to the remaining regions of the brain -- which do not participate in the motor and visual modules -- as comprising a non-motor, non-visual set. These areas do not tend to show consistent allegiances to any putative functional modules, indicating that their BOLD time courses are significantly different than those characterizing the motor and visual modules, and significantly different from one another. We observe that the regions in this set tend to have split allegiances: a pair of regions tends to have a 29\% probability of belonging to the same community, which is much lower than that observed in the visual (80\%) and motor (65\%) modules. We can interpret this difference by noting that the visual and motor modules contain areas required for task execution while the non-motor, non-visual set includes areas required for higher order cognitive processes including attention, executive function, and cognitive control. This suggests that cortices relevant for cognitive processes might only be transiently recruited, while cortices relevant for task execution are instead recruited consistently throughout learning.

\subsection{Dynamic Architecture of Learning}

\begin{figure}[h]
\centerline{\includegraphics[width=1\textwidth]{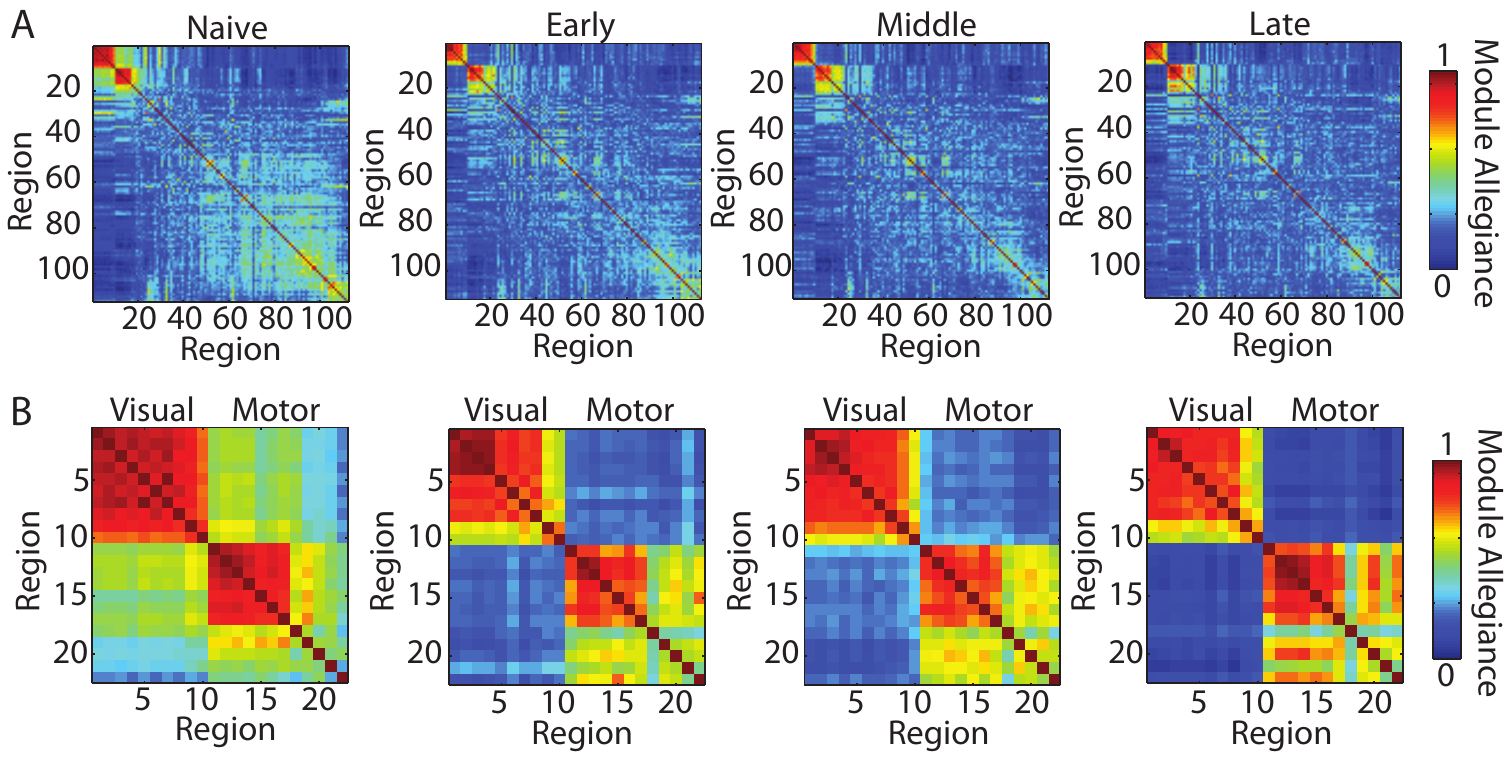}}
\caption{\textbf{Dynamic Brain Architecture Associated with Task Practice} \emph{(A)} In naive, early, middle, and late learning, the motor and visual modules evident in the stable architecture (Fig.~\ref{fig:stable}A) are also present. We observe a decrease in the strength of allegiance between regions in the non-motor and non-visual set. \emph{(B)} A blow up of the motor and visual module portions of panel \emph{(A)}, demonstrating the decrease in the strength of allegiance between these modules as learning progresses. }\label{fig:mods}
\end{figure}

We next sought to address the question: ``Do these modules or their interactions change with learning?'' To answer this question, we constructed separate module allegiance matrices for each of the four scanning sessions that were performed at approximately weeks 0, 2, 4, and 6 of the experiment (see Methods). We observed that the motor and visual modules evident in the summary network architecture are also consistently present across naive, early, middle, and late learning (Fig.~\ref{fig:mods}). However, the strength of interaction between these two modules appears to drastically decrease with task practice (see the transformation from warm to cool colors in the first and third quadrants of the matrices depicted in Fig.~\ref{fig:mods}B). We also observed that the strength of the module allegiance in the non-motor, non-visual set decreased with task practice, suggesting that these other brain areas are recruited to a lesser degree in late learning than in early learning. (See the Supplementary Information for a description of the robustness of these observations to several methodological choices.)

To quantify these observations, we estimated the recruitment of and integration between the three groups of brain areas defined from the module allegiance matrix summarizing task-based network architecture (see previous section): the motor module, the visual module, and the non-motor, non-visual set. Let $\mathcal{C}=\{C_1, \cdots, C_k\}$ be the partition of brain regions into these three groups. Then
\begin{equation}
I_{k_1, k_2} = \frac{\sum_{i\in C_{k_1}, j\in C_{k_2}}P_{ij}}{|C_{k_1}||C_{k_2}|}
\end{equation}
is the interaction strength between group $C_{k_1}$ and group $C_{k_2}$, where $|C_k|$ is the number of nodes in group $C_k$. To compute the average \emph{recruitment} of a single group to the task, we let $k_1 = k_2$. To compute the average \emph{integration} between two different groups ($k_1\neq k_2$), we calculate the normalized interaction strength
\begin{equation}
RI_{k_1, k_2} = \frac{I_{k_1, k_2}}{\sqrt{I_{k_1, k_1} I_{k_2, k_2}}}
\end{equation}
which accounts for statistical differences in group strength. We refer to both recruitment and integration more generally as \emph{brain network diagnostics}.

\begin{figure} [h]
\centerline{\includegraphics[width=.65\textwidth]{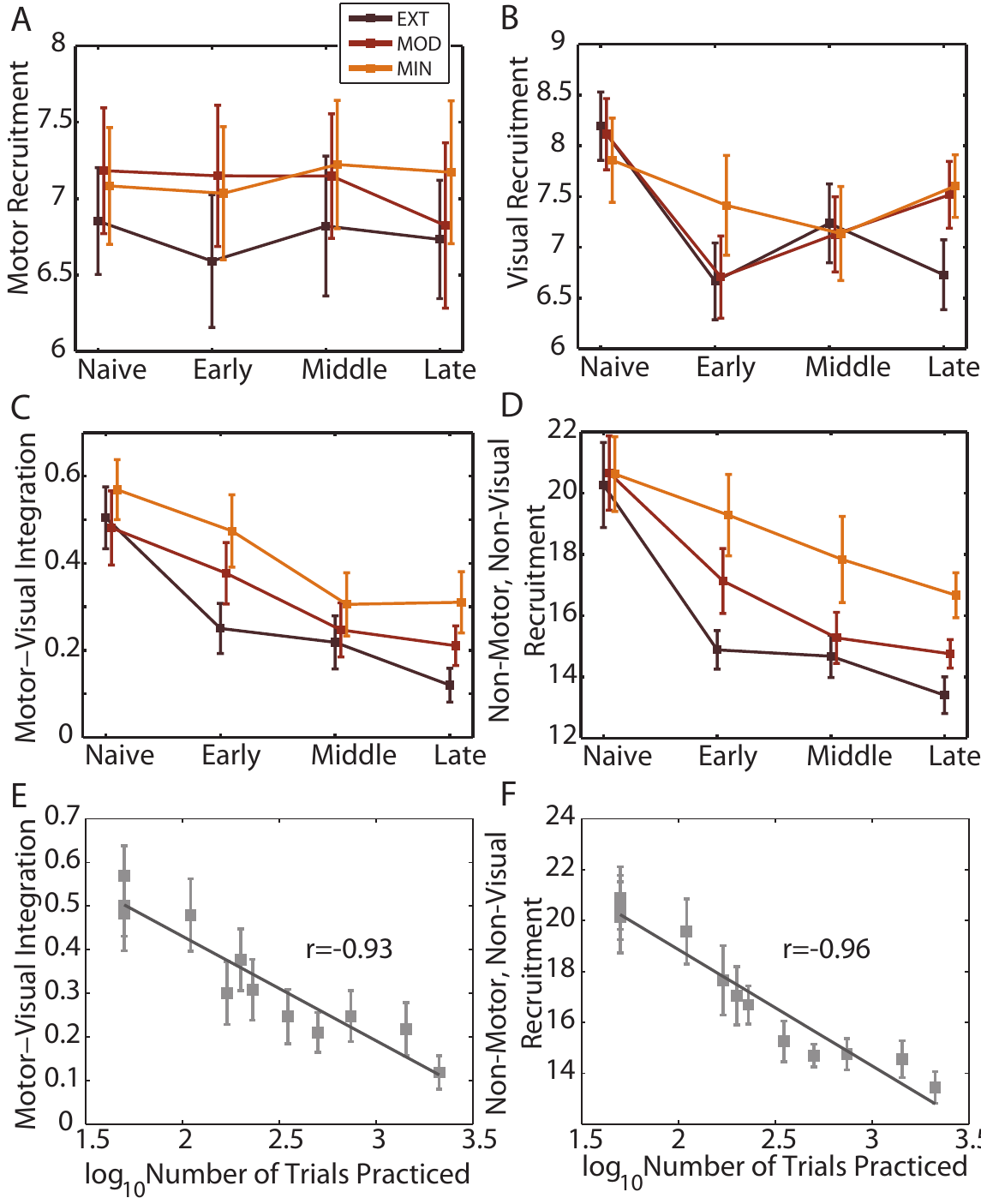}}
\caption{\textbf{Recruitment and Integration Modulated by Training} \emph{(A)} Motor and \emph{(B)} visual recruitment is unaffected by training intensity (extensively (maroon), moderately (red), and minimally (orange) trained sequences) and duration (naive, early, middle, and late). \emph{(C)} Integration between motor and visual modules and \emph{(D)} recruitment between non-motor and non-visual cortices decreases with training intensity and duration. \emph{(E,F)} The observations in panels \emph{(C,D)} that recruitment and integration depend on training intensity and duration can be parsimoniously described by a single latent variable: the number of trials practiced (i.e., ``depth''). Solid lines indicate best linear fit, and $r$ values indicate Pearson correlation coefficients. For the relationship between training duration, intensity, and depth, see Table~\ref{Tab2}. Error bars indicate standard deviation of the mean over participants. }\label{fig:int}
\end{figure}

Recruitment and integration in the motor and visual modules and the non-motor non-visual set are differentially modulated by training. Motor and visual modules are recruited steadily across task practice (Fig.~\ref{fig:int}A-B), consistent with their constant role in sustaining the performance of visually instructed motor sequences. In contrast, the integration between motor and visual modules decreases with both training intensity (i.e., extensive, moderate, or minimal) and training duration (i.e., naive, early, middle, and late). This decaying motor-visual integration suggests that motor and visual systems become more autonomous from one another with training (Fig.~\ref{fig:int}C). As motor and visual autonomy increases, the recruitment of areas in the non-motor, non-visual set also decreases (Fig.~\ref{fig:int}D), suggesting that task performance in later learning does not require higher order association areas to coordinate their function with one another.

The roles of training intensity and duration on brain network diagnostics (Fig.~\ref{fig:int}C-D) can be more parsimoniously described by training depth, as defined by the number of trials practiced (Fig.~\ref{fig:int}E-F). To quantify these observations, we define the \emph{training-dependent modulation} of brain network diagnostics as $r-1$, where $r$ is the Pearson correlation coefficient between the logarithm of the number of trials practiced and the diagnostic value. Motor-visual integration showed significant training-dependent modulation ($r=-0.93$, $p= 5.66 \times 10^{-6}$), as did non-motor, non-visual recruitment ($r=-0.96$, $p=4.32 \times 10^{-7}$).

\subsection{System-Level Correlates of Performance and Learning}

We next ask the question: ``Could the interplay between task-relevant modules or the involvement of association cortex explain the dramatic differences between individual people in their capacity to learn?''. To answer this question, we performed a more fine-grained analysis at the level of single individuals.

\begin{figure} [h]
\centerline{\includegraphics[width=.65\textwidth]{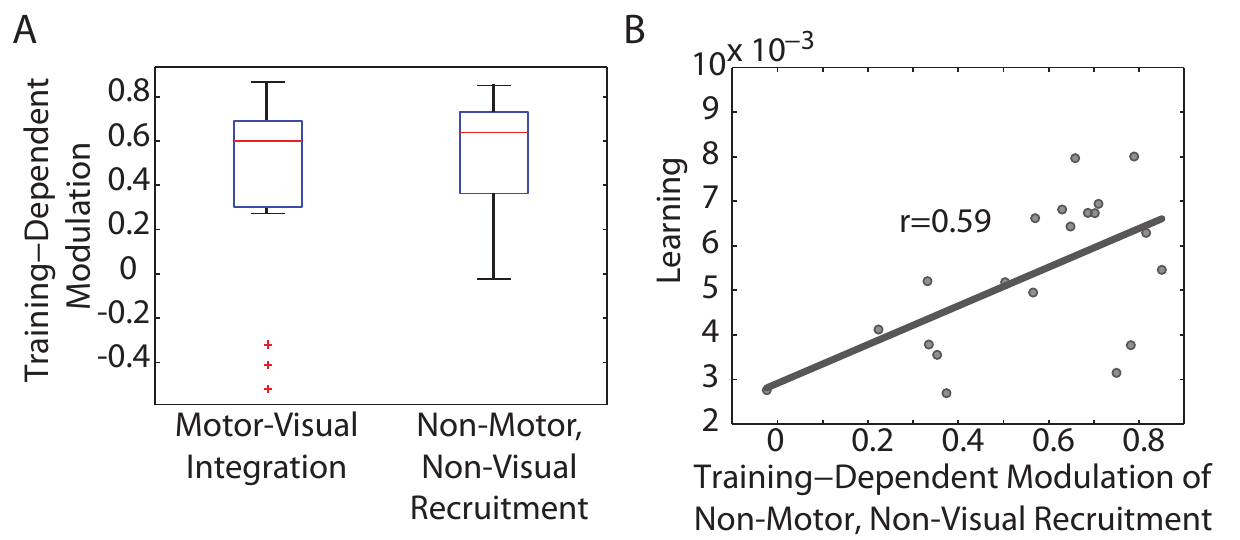}}
\caption{\textbf{Individual Differences in Brain Network Architecture Map to Task Performance and Task Learning.} \emph{(A)} Boxplots of task-dependent modulation (correlation between network diagnostic and number of trials practiced) for both motor-visual integration (left) and non-motor, non-visual recruitment (right). \emph{(B)} Scatter plot of learning and training-dependent modulation of non-motor, non-visual recruitment ($r=0.59$, $p=0.0062$).  }\label{fig:beh}
\end{figure}

We observed that motor-visual integration and non-motor, non-visual recruitment display training-dependent modulation at the single subject level (Fig.~\ref{fig:beh}A). However, the strength of this modulation varied over individuals, indicating that some individuals showed a stronger growth in sensorimotor autonomy than others. We hypothesized that individuals whose brain networks displayed greater sensorimotor autonomy during training would learn better than those whose brain networks maintained strong motor-visual integration and non-motor non-visual recruitment in late learning.

To test this hypothesis, we examined the relationship between a statistic summarizing learning in individual subjects and the strength of training-dependent modulation. To calculate the learning parameter, we defined the \emph{movement time} to be the duration from the first to last button press for a given sequence. We then defined a statistic of learning rate to be the exponential drop-off parameter of the movement times \cite{Bassett2011b}, collated from home training sessions over the course of the 6-week experiment. We did not observe a significant relationship between training-dependent modulation of motor-visual integration and learning rate ($r=0.42$, $p=0.0650$). This suggests that the separation between motor and visual modules is driven by task practice (which is common across the group) rather than learning rate (which differs across participants). Conversely, we observed a strong correlation between training-dependent modulation of non-motor non-visual recruitment and learning rate ($r=0.59$, $p=0.0062$; Fig.~\ref{fig:beh}B). This suggests that individuals who are able to disband extraneous brain areas in the non-motor, non-visual set over the course of training are better able to learn than individuals who maintain these extraneous areas online through later learning.

\subsection{A Network Driver of Individual Differences in Learning}

Our observation that the training-dependent modulation of non-motor, non-visual areas is correlated with individual differences in learning is based on system-scale measurements. Here we ask which specific elements -- or functional connections -- within that system drive the observed prediction.

To address this question, we calculate the training-induced modulation of each edge connecting pairs of brain areas in the non-motor, non-visual set for each participant and we ask whether this value is correlated with the learning rate parameter. We collate the set of edges for which this correlation is significant ($p<0.05$ uncorrected; Pearson's $r$) and refer to this set of edges collectively as the \emph{predictive network}.

\begin{figure} [h]
\centerline{\includegraphics[width=.65\textwidth]{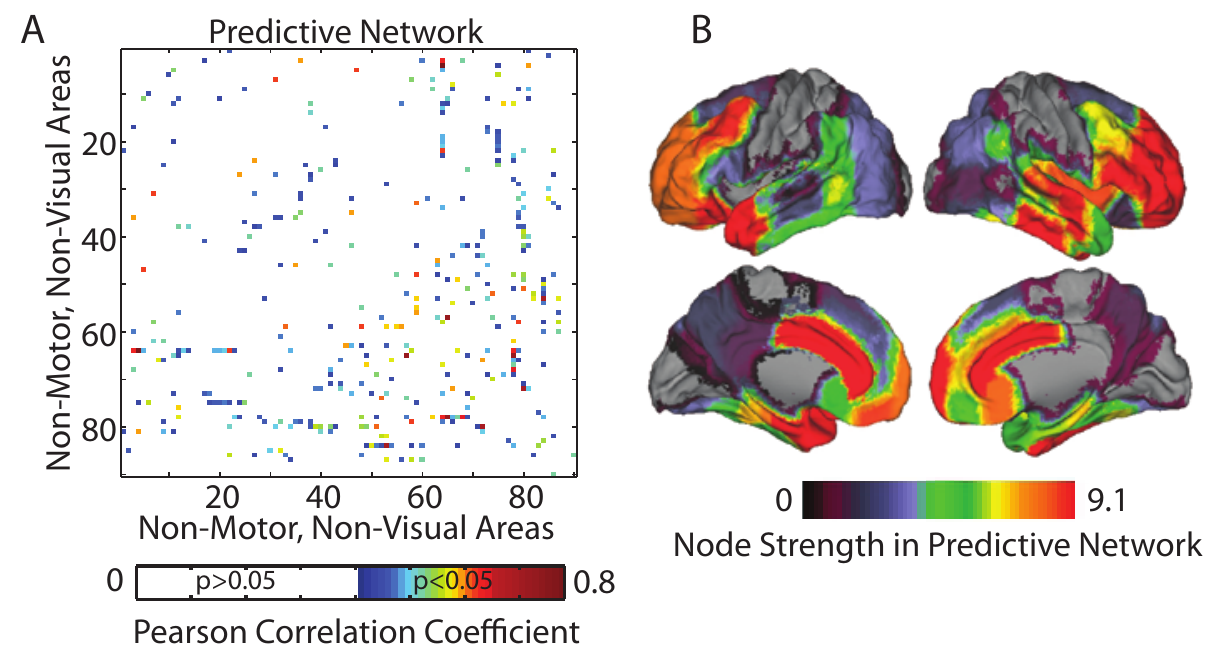}}
\caption{\textbf{The Release of a Frontal-Cingulate Control Network Predicts Individual Differences in Learning.} \emph{(A)} Elements in the predictive network are given by the statistically significant ($p<0.05$ uncorrected) Pearson correlation coefficients between individual differences in training-induced modulation and individual differences in learning. Color indicates the magnitude of the Pearson correlation coefficient. \emph{(B)} The strength of brain areas mapped on to the cortical surface using Caret \cite{VanEssen2001}. The strength of area $i$ is given by the sum of column $i$ in the predictive network. Warm colors indicate high strength in the predictive network and cool colors indicate low strength in the predictive network. }\label{fig:pred}
\end{figure}

The predictive network, which spans approximately 96\% of the non-motor non-visual set, is composed of 180 functional connections that drive individual differences in learning (Fig.~\ref{fig:pred}A). These connections are distributed asymmetrically throughout the network: a few brain areas boast many connections in the predictive network, while most brain areas boast only a few. This observation can be quantified by the regional \emph{strength}, defined as the sum of the weights of the edges emanating from that area \footnote{Recall that the weights of the predictive network are given by the Pearson correlation coefficient $r$ between the training-induced modulation of that edge and the learning parameter.}. Consistent with our qualitative observations, the distribution of strength values over brain areas is highly skewed ($s=1.89$; Fig.~\ref{fig:pred}B), significantly more so than expected in a random network null model (non-parametric test $p=0.00009$; see SI for statistical details). We observe that areas of high strength in the predictive network are predominantly located in frontal and cingulate cortices (Fig.~\ref{fig:pred}C), suggesting that the training-induced release of a frontal-cingulate system predicts individual differences in learning.

\section*{Discussion}

Here we address the hypothesis that long-term skill acquisition requires changes in the recruitment of and integration between functional systems \cite{Dayan2011}. We acquired fMRI data during the performance of a motor sequence task over 6 weeks of practice. We used novel network analysis algorithms to map data-derived functional modules of synchronized brain areas to anatomical boundaries, quantify the recruitment of each module separately, and calculate the functional integration between modules as a function of training intensity. We observed that motor areas and primary visual areas form two separate functionally cohesive modules whose recruitment does not change with training but whose integration with one another decreases as sequence performance becomes more automatic. We also observe that the remaining brain areas (which we refer to as the non-motor, non-visual set) display decreasing integration as training progresses, and individual differences in this decrease -- particularly in the fronto-cingulate system -- predict individual differences in learning.

\subsection{Learning-Induced Changes in Sensorimotor Systems}

How sensorimotor areas change their activity during motor skill learning has been the topic of extensive study over recent years \cite{Dayan2011}. Collectively, these studies demonstrate that in early (fast) learning, sequential motor tasks (i) decrease the magnitude of the task-evoked BOLD response in the dorsolateral prefrontal cortex (DLPFC), primary motor cortex (M1), and presupplementary motor area (preSMA) \cite{Floyer2005,Sakai1999}, and (ii) increase the magnitude of the task-evoked BOLD response in the premotor cortex, supplementary motor area (SMA), parietal regions, striatum, and cerebellum \cite{Grafton2002,Honda1998,Floyer2005}. Meanwhile, long term sequence learning is linked to (i) increases in the magnitude of the task-evoked BOLD response in M1 \cite{Floyer2005}, primary somatosensory cortex \cite{Floyer2005}, SMA \cite{Lehericy2005}, and putamen \cite{Lehericy2005,Floyer2005}, and (ii) decreases in the magnitude of the task-evoked BOLD response in lobule VI of the cerebellum \cite{Lehericy2005}. For complementary results in our data, see the SI.

In contrast to these previous studies, we uncover network adaptations across a continuum of learning (see Table~\ref{Tab2}), rather than examining early or late learning alone. Furthermore, we probe the dynamic, learning-induced integration and release of distributed cognitive systems, processes that are inaccessible to the study of task-related activity alone. Activity and functional connectivity provide two different windows into changes in brain function, as evidenced by recent work in learning \cite{Bassett2011b}, memory \cite{Sieben2013b}, disease diagnosis \cite{Bassett2012a}, intervention \cite{Patel2013}, and genetics \cite{Esslinger2009}. As a complement to studies of task-related activity, functional connectivity has the added advantage of providing information on the interregional interactions \cite{Sun2007}, that together form the structure of cognitive models of learning \cite{Dayan2011}.

Our results suggest that motor and visual systems transition from being heavily integrated early to becoming autonomous units later during learning, such that each system is performing independent computations hallmarked by unique temporal profiles of BOLD activity.  This growing autonomy is consistent with the functional requirements of skill acquisition. In early learning, subjects are required to master multiple tasks that necessitate the integration of vision and motion: the use of a response box, decoding of the visual stimulus, performance of precise movements, balancing of attention between visual stimuli, and switching between different sequences of movements. In addition, until they have developed an internal model or representation of each 10-element sequence, they are dependent on each visual cue to guide subsequent key presses. Once a sequence is well learned, the only visual information that is needed is the initial cue indicating which sequence to perform. At this point, subjects execute an entire sequence in an extremely rapid predictive manner, and without the reliance on visual instruction from individual key press stimuli.

Furthermore, the growth of sensorimotor autonomy is consistent with a \emph{neural efficiency hypothesis}. Such a hypothesis could suggest that as learning progresses, the cognitive resources that are initially devoted early in learning are no longer needed. Instead, the cortical system will tend to economize resources \cite{Kelly2005,Petersen1998} and limit unnecessary communication and transmission to enable automaticity.

\subsection{A Frontal-Cingulate Network Predicts Individual Differences in Learning}

As motor and visual systems increase in autonomy, the remaining areas of the brain decrease in recruitment. Individual differences in the recruitment of this non-motor non-visual network was correlated with individual differences in learning: individuals who were able to disband this network during task practice learned better than those who did not. Such a correlation was not observed between learning and motor-visual recruitment or integration. These results are consistent with those of a prior study linking the BOLD amplitude of motor and premotor cortices with task performance but not with learning \cite{Orban2010}. Thus, while motor-visual systems are required for subject-general processes such as task execution, the non-motor non-visual network encapsulates subject-specific processes including individual differences in learning.

We further identified the distributed network of individual functional connections whose training-dependent release predicted individual differences in learning. These connections emanate predominantly from frontal and anterior cingulate cortices, two hubs of known cognitive control systems: the frontal-parietal and the cingulo-opercular systems. These two systems are characterized by different functional connectivity patterns at rest \cite{Power2011} and are thought to support distinct functional roles \cite{Elton2013}: task-switching \cite{Stoet2009} and task-set maintenance \cite{Shenhav2013}, respectively. This top down level of control varies depending on the organism's goals and the characteristics of the given task \cite{Chrysikou2013}. While too much cognitive control can impede learning \cite{ThompsonSchill2009}, at appropriate levels of engagement, the two processes interact symbiotically by modulating common anatomical structures (prefrontal cortex and basal ganglia) \cite{Collins2013}. It is intuitively plausible that cognitive control is particularly critical during early skill acquisition \cite{Hikosaka2002,Petersen1998} and becomes less so as skills reach automaticity. Such an argument is supported by our results demonstrating that these two hubs of known cognitive control systems become disengaged from the rest of the network through the 6 weeks of training.

Importantly, the neurophysiological processes that we have described are extracted from 4 scanning sessions held every 1.5--2 weeks encompassing a 6 week period of task practice. The behavioral estimates of learning, on the other hand, are extracted from home training sessions taking place on the days between scans. The temporal separation of the data from the scanning sessions (which we used to estimate recruitment and integration) and the home training (which we used to estimate learning) ensures that these results are predictive rather than simply correlative. Prior reports of neurophysiological predictors of learning have either focused on gross network characteristics \cite{Bassett2011b,Bassett2013a} or isolated functional connections \cite{Buchel1999}, thereby inhibiting interpretations at the level of dynamically integrated cognitive systems. Here we employ a new set of methodological approaches that enables us to identify the cognitive networks engaged during the task, track changes in the recruitment of and integration between networks during task practice, extract a predictive network of functional interactions that drive individual differences in learning, and confirm the statistical significance of this predictive network using non-parametric null models. Collectively, these approaches enable us to link characteristics of network adaptivity to known cognitive systems and the broader cognitive neuroscience literature describing their functions.

\subsection{Methodological Considerations}

In choosing and developing the approaches utilized in this work, we have considered several factors. First, we have chosen to perform a whole-brain analysis, rather than examining a handful of regions displaying task-related BOLD activation, to enhance our ability to detect features of inter-regional communication patterns that either directly or indirectly support task performance and motor learning. Second, we have chosen to examine functional connectivity rather than BOLD activity to enable us to probe the recruitment of and integration between cognitive systems engaged in motor skill acquisition over both early and late learning time periods. Third, we have chosen to focus on network modularity and the changes in putative functional modules during skill learning rather than on more global or local graph metrics which are less interpretable in the framework of known cognitive systems. Finally, we have chosen to present and characterize module allegiance matrices rather than individual partitions because they more accurately display the inherent overlaps between putative functional modules, which -- rather than forming independent systems -- can display behaviorally relevant interactions that can differ across individuals and can change over the time scales of learning.

Several methodological choices deserve additional consideration. First, region size has been shown to have a significant effect on structural network architecture estimated from other imaging modalities. In the SI, we show that region size does not account for the observed structure in module allegiance matrices. Second, two common approaches to network analysis include (i) the examination of the fully weighted network and (ii) the examination of a weighted network that has been thresholded using some test for the statistical significance of individual edges in the network \cite{Bassett2013b}. In the SI, we demonstrate that our results are robustly observed in both thresholded and unthresholded module allegiance matrices. Third, in the introduction of any new method, it is important to ask whether similar results could have been uncovered using a simpler approach. In the SI, we demonstrate that module allegiance matrices provide a level of sensitivity to learning-related changes in brain network architecture that is not observed in functional connectivity matrices alone. Because the module allegiance matrix represents only information about the network topology, it is unaffected by measurement noise that may change the mean or distribution of functional connectivity values.

\subsection{Broader Implications for Cognitive and Clinical Neuroscience}

Our results highlight several important opportunities for the cognitive neuroscience of learning. The traditional mapping of single functions to single brain areas is likely an overly simplistic account of cognitive function, unable to accurately depict its true complexity \cite{Fedorenko2014}. Network science provides a complement to traditional univariate contrast analyses, by providing access to neurophysiological processes that would otherwise be hidden. These processes include changes in BOLD activity and connectivity that occur over the time scales of skill acquisition. Indeed, our results suggest that single brain areas are unlikely to map to single cognitive functions in all task conditions and at all times. Instead, brain regions may alter their allegiance to putative functional modules across time and in different task states according to the most relevant association for the cognitive process at hand \cite{Bassett2013a}. We speculate that the further development of empirical, computational, and theoretical methods to probe these intricacies are likely to be of increasing relevance in addressing the challenging questions that currently face cognitive neuroscience.

Our findings in a healthy adult cohort could also inform our understanding of and hypotheses regarding disease and injury-induced changes in the abnormal brain. Neurodegeneration, movement disorders, and stroke can be associated with a large-scale reorganization of the motor system \cite{Frey2011}. In these scenarios, both individual circuits and their interactions with one another can be altered \cite{Beeler2013}. Our results suggest that the anatomical (in motor, visual, or cognitive control systems) and topological (within versus between cognitive and sensorimotor systems) location of these alterations account for a significant amount of variance in individual differences in response to rehabilitation and treatment \cite{Cumberland2009}, informing therapeutic manipulations to circuits in the form of non-invasive brain stimulation \cite{Sandrini2013}.

\subsection{Conclusions}

The dynamic integration of distributed neural circuits necessary to transform a motor skill from slow and challenging to fast and automatic has evaded description due to statistical and mathematical limitations in current analysis frameworks. Here we utilize dynamic network neuroscience approaches to expose a learning-induced autonomy of sensorimotor systems and to uncover a distributed network of frontal and anterior cingulate cortices whose disengagement predicts individual differences in learning. These results provide a cohesive and statistically principled account of the dynamics of distributed and integrated circuits during cognitive processes underlying skill learning in humans.

\begin{figure}[h]
\centerline{\includegraphics[width=0.65\textwidth]{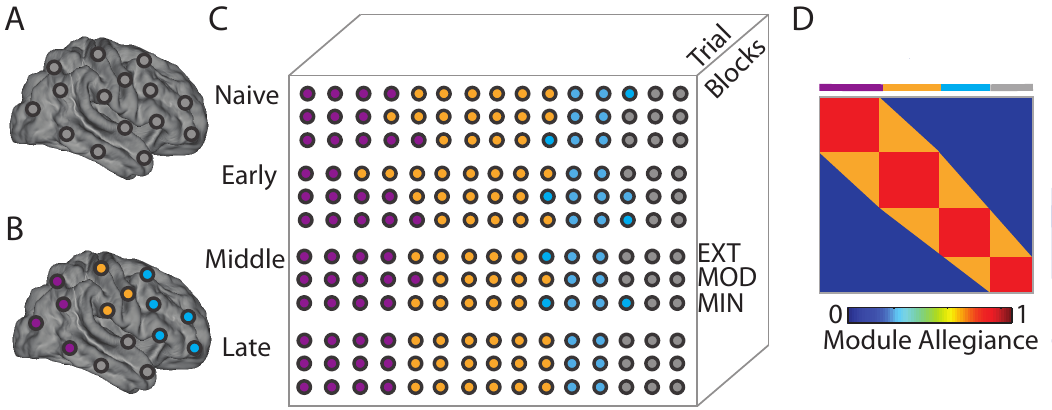}}
\caption{\textbf{Schematic of Methods} \emph{(A)} We parcellate the brain into 112 cortical and subcortical regions based on the Harvard-Oxford atlas.  \emph{(B)} We calculate the functional connectivity between these regions to create a functional network, and we cluster regions within the functional network using community detection techniques. \emph{(C)} We collate the community assignments (\emph{partitions}) across different time scales of learning (naive, early, middle, and late), different depths of training (EXT, MOD, MIN), and different subjects. \emph{(D)} We create a module allegiance matrix whose elements indicate the probability that any two regions are classified into the same network community. }\label{fig:methods}
\end{figure}

\bigskip
\begin{methods} 

Twenty healthy, right-handed participants (11 females and 9 males; mean age 24) volunteered with informed consent in accordance with the Institutional Review Board/Human Subjects Committee, University of California, Santa Barbara. Participants completed a minimum of 30 home training sessions as well as 4 fMRI sessions over the course of approximately 6 weeks. Each participant practiced a set of 10-element, visually-presented finger-movement sequences with his/her right hand.

Participants practiced 6 motor sequences (50 trials/sequence) while lying in an fMRI scanner on day 1 of the experiment (this state is hereafter termed ``naive''), two weeks after day 1 (``early''), 4 weeks after day 1 (``middle''), and 6 weeks after day 1 (``late''). In between scanning sessions, participants practiced the same 6 sequences at home on their laptop computers. Two sequences were practiced extensively (``EXT'', 75 trials/sequence/session), two sequences were practiced moderately (``MOD'', 10 trials/sequence/session), and two sequences were practiced minimally (``MIN'', 1 trials/sequence/session).

Each scanning session was comprised of 6-10 experiment blocks. For each subject and each block, we calculated the functional connectivity between all possible pairs of $N=112$ cortical and subcortical regions defined by the Harvard-Oxford atlas (Fig.~\ref{fig:methods}A) using a mean square coherence of the wavelet scale 2 time series (approximately corresponding to fluctuations in the BOLD signal in the frequency band 0.06-0.12 Hz). For each coherence matrix (or \emph{functional network}), we performed 100 optimizations of the modularity quality function \cite{Newman2006} to partition brain regions into network communities. Each community is composed of regions with high intra-community coherence, and therefore corresponds to a putative functional module (Fig.~\ref{fig:methods}B). For statistical robustness, we extracted a consensus partition from these optimizations by comparison to a non-parametric null model \cite{Bassett2012b}. This set of procedures provides us with a partition of brain regions into communities for each subject, scanning session, sequence type, and trial blocks.

Brain-surface visualizations utilize Caret software (\url{brainvis.wustl.edu}).

See the SI for additional information.

\end{methods}


\newpage
\begin{addendum}
 \item  DSB acknowledges support from the Alfred P. Sloan Foundation, the Army Research Laboratory, and the Institute for Translational Medicine and Therapeautics. MY acknowledges support from the Applied Mathematics and Computational Science Graduate Program at the University of Pennsylvania. NFW and STG were supported by PHS Grant NS44393 and the Institute for Collaborative Biotechnologies through grant W911NF-09-0001 from the U.S. Army Research Office. The content is solely the responsibility of the authors and does not necessarily represent the official views of any of the funding agencies. We thank David Baker, Sarah Feldt Muldoon, and Qawi Telesford for comments on an earlier version of the manuscript.
  \item [Author Contributions] D.S.B. formulated the project; N.F.W. and S.T.G. performed the experiments; D.S.B., N.F.W., and M.Y. did the computations; and D.S.B., M.Y., N.F.W., and S.T.G. wrote the manuscript.
  \item [Competing Interests] The authors declare that they have no competing financial interests.
  \item [Correspondence] Correspondence and requests for materials should be addressed to D.S.B.~(e-mail: \url{dsb@seas.upenn.edu}).
\end{addendum}

\clearpage
\newpage

\begin{table}
\hfill{}
\begin{tabular}{|l|l|}
\hline
\small{Motor} & \small{Visual}\\
\hline
\small{L,R Precentral gyrus}              & \small{L,R Intracalcarine cortex}\\
\small{L,R Postcentral gyrus}             & \small{L,R Cuneus cortex}\\
\small{L,R Superior parietal lobule}      & \small{L,R Lingual gyrus}\\
\small{L,R Supramarginal gyrus, anterior} & \small{L,R Supercalcarine cortex}\\
\small{L,R Supplemental motor area}       & \small{L,R Occipital Pole}\\
\small{L~~ Parietal operculum cortex}       & \\
\small{R~~ Supramarginal gyrus, posterior}  & \\
\hline
\end{tabular}
\hfill{}
\caption{\textbf{Brain Areas in Motor and Visual Modules}}
\label{Tab1}
\end{table}

\begin{center}
\begin{table}
{\normalsize
\hfill{}
\begin{tabular}{|c||c|c|c|c|}
\hline
~ &                     Naive    & Early   & Middle  & Late \\
\hline
MIN Sequences        & 50     & 110         & 170        & 230          \\
MOD Sequences        & 50     & 200         & 350        & 500          \\
EXT Sequences        & 50     & 740         & 1430       & 2120         \\
\hline
\end{tabular}}
\hfill{}
\caption{\textbf{Relationship Between Training Duration, Intensity, and Depth}. The number of trials (i.e., ``depth'') of each sequence type (i.e., ``intensity'') completed after each scanning session (i.e., ``duration'') averaged over the 20 participants.
\label{Tab2}
}
\end{table}
\end{center}

\clearpage
\newpage
\bibliography{bibfile}

\end{document}


\maketitle

\begin{affiliations}
 \item Complex Systems Group, Department of Bioengineering, University of Pennsylvania, Philadelphia, PA, 19104, USA
 \item Department of Electrical Engineering, University of Pennsylvania, Philadelphia, PA, 19104, USA
 \item Applied Mathematics and Computational Science Graduate Group, University of Pennsylvania, Philadelphia, PA, 19104, USA
 \item Department of Psychological and Brain Sciences and UCSB Brain Imaging Center, University of California, Santa Barbara, CA 93106, USA
\end{affiliations}

\newpage

\tableofcontents
\listoffigures
\listoftables
\newpage

\section*{Materials and Methods}
\addcontentsline{toc}{section}{Materials and Methods}

\subsection{Experiment and Data Acquisition}
\addcontentsline{toc}{subsection}{Experiment and Data Acquisition}

\subsubsection*{Ethics Statement}
\addcontentsline{toc}{subsubsection}{Ethics Statement}

Twenty-two right-handed participants (13 females and 9 males; the mean age was about 24) volunteered with informed consent in accordance with the Institutional Review Board/Human Subjects Committee, University of California, Santa Barbara.


\subsubsection*{Experiment Setup and Procedure}
\addcontentsline{toc}{subsubsection}{Experiment Setup and Procedure}

We excluded two participants from the investigation: one participant failed to complete the experiment, and the other had excessive head motion. Our investigation therefore includes twenty participants, who all had normal/corrected vision and no history of neurological disease or psychiatric disorders. Each of these participants completed a minimum of 30 behavioral training sessions as well as 3 fMRI test sessions and a pre-training fMRI session. Training began immediately following the initial pre-training scan session. Test sessions occurred after every 2-week period of behavioral training, during which at least 10 training sessions were required. The training was done on personal laptop computers using a training module that was installed by the experimenter (N.F.W.). Participants were given instructions for how to run the module, which they were required to do for a minimum of 10 out of 14 days in a 2-week period. Participants were scanned on the first day of the experiment (scan 1), and then a second time approximately 14 days later (scan 2), once again approximately 14 days later (scan 3), and finally 14 days after that (scan 4).

\begin{figure}
\begin{center}
\includegraphics[width=0.7\linewidth]{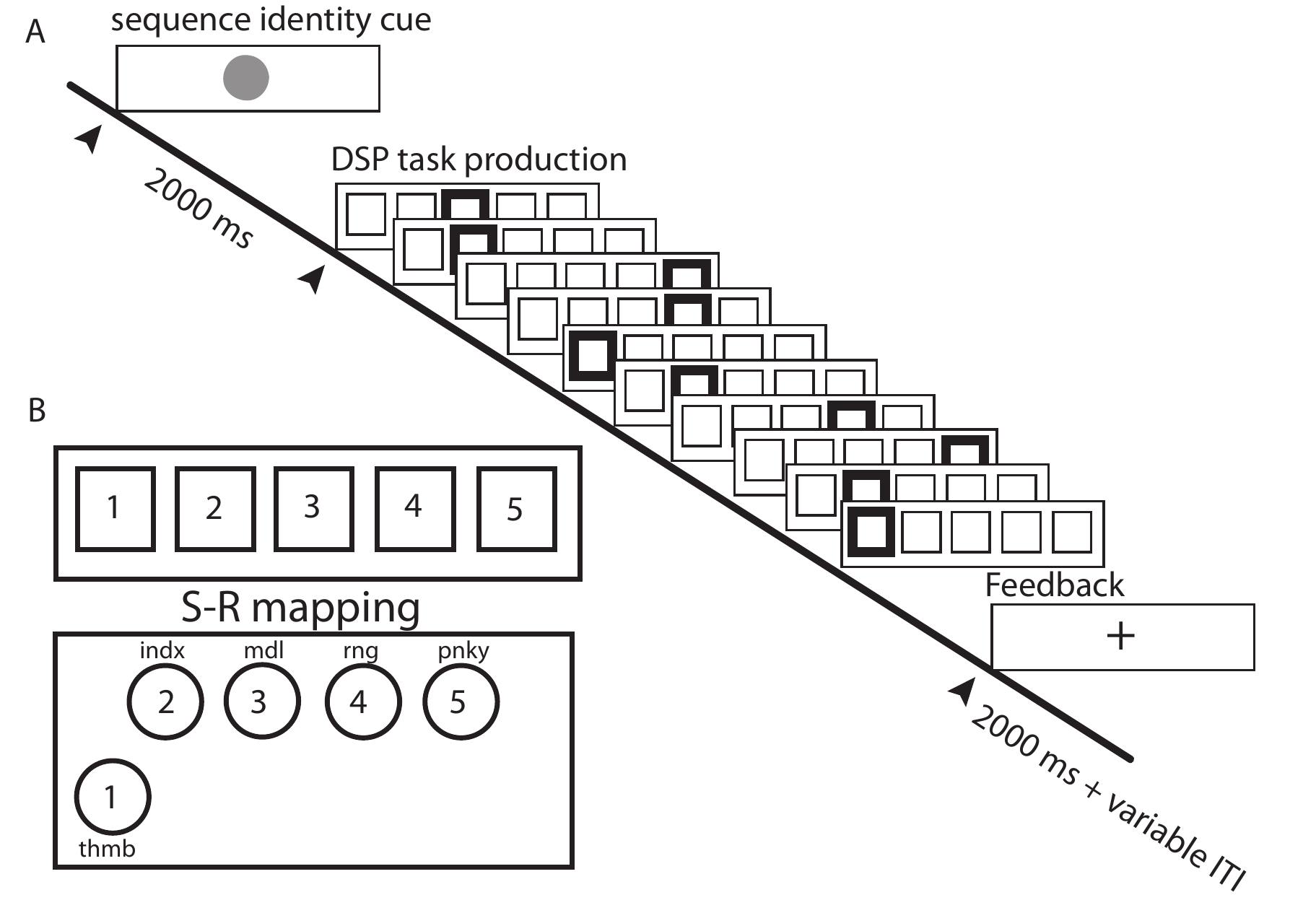}
    \caption[Trial Structure and Stimulus-Response (S-R) Mapping]{\label{fig:dsp} \textbf{Trial Structure and Stimulus-Response (S-R) Mapping} \emph{(A)} Each trial began with the presentation of a sequence identity cue that remained on screen for $2$ seconds. Each of the $6$ trained sequences was paired with a unique identity cue. A discrete sequence-production (DSP) event structure was used to guide sequence production. The onset of the initial DSP stimulus (thick square, colored red in the task) served as the imperative to produce the sequence. A correct key press led to the immediate presentation of the next DSP stimulus (and so on) until the $10$-element sequence was correctly executed. Participants received a feedback `+' to signal that a sequence was completed and to wait (approximately $0$--$6$ seconds) for the start of the next trial. This waiting period is called the `inter-trial interval' (ITI). At any point, if an incorrect key was hit, a participant would receive an error signal (not shown in the figure) and the DSP sequence would pause until the correct response was received. \emph{(B)} There was a direct S-R mapping between a conventional keyboard or an MRI-compatible button box (see the lower left of the figure) and a participant's right hand, so the leftmost DSP stimulus cued the thumb and the rightmost stimulus cued the pinky finger. Note that the button location for the thumb was positioned to the lower left to achieve maximum comfort and ease of motion.
    }
    \end{center}
\end{figure}

We asked participants to practice a set of 10-element sequences that were presented visually using a discrete sequence-production (DSP) task by generating responses to sequentially presented stimuli (see Fig.~\ref{fig:dsp}) using a laptop keyboard with their right hand. Sequences were presented using a horizontal array of 5 square stimuli; the responses were mapped from left to right, such that the thumb corresponded to the leftmost stimulus and the smallest finger corresponded to the rightmost stimulus. A square highlighted in red served as the imperative to respond, and the next square in the sequence was highlighted immediately following each correct key press. If an incorrect key was pressed, the sequence was paused at the error and was restarted upon the generation of the appropriate key press.

\begin{figure}
\includegraphics[width=0.7\linewidth]{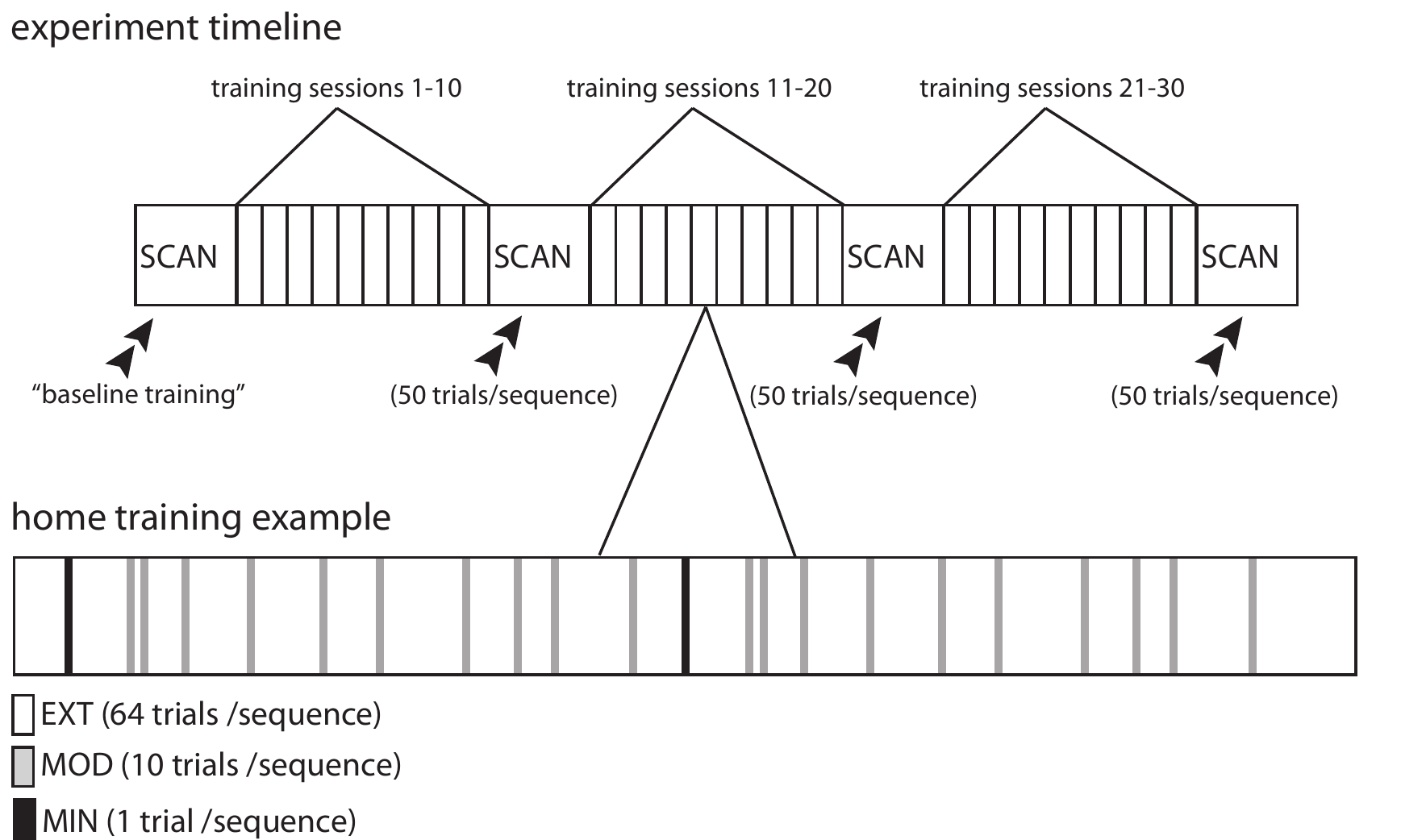}
    \caption[Experiment Timeline]{\label{fig:exptime} \textbf{Experiment Timeline} Training sessions in the MRI scanner during the collection of blood-oxygen-level-dependent (BOLD) signals were interleaved with training sessions at home. Participants first practiced the sequences in the MRI scanner during a baseline training session \emph{(top)}. Following every approximately 10 training sessions (see Supplementary Table 1), participants returned for another scanning session. During each scanning session, a participant practiced each sequence for 50 trials. Participants trained at home between the scanning sessions \emph{(bottom)}. During each home training session, participants practiced the sequences in a random order. (We determined a random order using the Mersenne Twister algorithm of Nishimura and Matsumoto \cite{Matsumoto1998} as implemented in the random number generator {\tt rand.m} of MATLAB version 7.1). Each EXT sequence was practiced for 64 trials, each MOD sequence was practiced for 10 trials, and each MIN sequence was practiced for 1 trial.
    }
\end{figure}

Participants had an unlimited amount of time to respond and to complete each trial. All participants trained on the same set of 6 different 10-element sequences, which were presented with 3 different levels of exposure. We organized sequences so that each stimulus location was presented twice and included neither stimulus repetition (e.g., ``11'' could not occur) nor regularities such as trills (e.g., ``121'') or runs (e.g., ``123''). Each training session (see Fig.~\ref{fig:exptime}) included 2 extensively trained sequences (``EXT'') that were each practiced for 64 trials, 2 moderately trained sequences (``MOD'') that were each practiced for 10 trials, and 2 minimally trained sequences (``MIN'') that were each practiced for 1 trial. (See Table S1 for details of the number of trials composed of extensively, moderately, and minimally trained sequences during home training sessions.) Each trial began with the presentation of a sequence-identity cue. The purpose of the identity cue was to inform the participant what sequence they were going to have to type.
For example, the EXT sequences were preceded by either a cyan (sequence A) or magenta (sequence B) circle.  Participants saw additional identity cues for the MOD sequences (red or green triangles) and for the MIN sequences (orange or white stars, each of which was outlined in black). No participant reported any difficulty viewing the different identity cues. Feedback was presented after every block of 10 trials; this feedback detailed the number of error-free sequences that the participant produced and the mean time it took to complete an error-free sequence.

Each fMRI test session was completed after approximately 10 home training sessions, and each participant participated in 3 test sessions. In addition, each participant had a pre-training scan session that was identical to the other test scan sessions immediately prior to the start of training (see Fig.~\ref{fig:exptime}). To familiarize participants with the task, we gave a brief introduction prior to the onset of the pre-training session. We showed the participants the mapping between the fingers and the DSP stimuli, and we explained the significance of the sequence-identity cues.

To help ease the transition between each participant's training environment and that of the scanner, padding was placed under his/her knees to maximize comfort. Participants made responses using a fiber-optic response box that was designed with a similar configuration of buttons as those found on the typical laptop used during training. See the lower left of Fig.~\ref{fig:dsp} for a sketch of the button box used in the experiments. For instance, the center-to-center spacing between the buttons on the top row was 20 mm (compared to 20 mm from ``G'' to ``H'' on a recent MacBook Pro), and the spacing between the top row and lower left ``thumb'' button was 32 mm (compared to 37 mm from ``G'' to the spacebar on a MacBook Pro). The response box was supported using a board whose position could be adjusted to accommodate a participant's reach and hand size. Additional padding was placed under the right forearm to minimize muscle strain when a participant performed the task. Head motion was minimized by inserting padded wedges between the participant and the head coil of the MRI scanner. The number of sequence trials performed during each scanning session was the same for all participants, except for two abbreviated sessions that resulted from technical problems. In each case that scanning was cut short, participants completed 4 out of the 5 scan runs for a given session. We included data from these abbreviated sessions in this study.

Participants were tested inside of the scanner with the same DSP task and the same 6 sequences that they performed during training. Participants were given an unlimited time to complete trials, though they were instructed to respond quickly but also to maintain accuracy. Trial completion was signified by the visual presentation of a fixation mark ``+'', which remained on the screen until the onset of the next sequence-identity cue. To acquire a sufficient number of events for each exposure type, all sequences were presented with the same frequency. Identical to training, trials were organized into blocks of 10 followed by performance feedback. Each block contained trials belonging to a single exposure type and included 5 trials for each sequence. Trials were separated by an inter-trial interval (ITI) that lasted between 0 and 6 seconds (not including any time remaining from the previous trial). Scan epochs contained 60 trials (i.e., 6 blocks) and consisted of 20 trials for each exposure type. Each test session contained 5 scan epochs, yielding a total of 300 trials and a variable number of brain scans depending on how quickly the task was performed.


\subsubsection*{Behavioral Apparatus}
\addcontentsline{toc}{subsubsection}{Behavioral Apparatus}

Stimulus presentation was controlled during training using a participant's laptop computer, which was running Octave 3.2.4 (an open-source program that is very similar to {\sc Matlab}) in conjunction with PsychtoolBox Version 3. We controlled test sessions using a laptop computer running {\sc Matlab} version 7.1 (Mathworks, Natick, MA). We collected key-press responses and response times using a custom fiber-optic button box and transducer connected via a serial port (button box: HHSC-$1\times4$-L; transducer: fORP932; Current Designs, Philadelphia, PA).


\subsubsection*{Behavioral Estimates of Learning}
\addcontentsline{toc}{subsubsection}{Behavioral Estimates of Learning}

Our goal was to study the relationship between brain network organization and learning. To ensure independence of these two variables, we extracted brain network structure during the 4 scanning sessions, and we extracted behavioral estimates of learning in home training sessions across the 6 weeks of practice.

For each sequence, we defined the movement time (MT) as the difference between the time of the first button press and the time of the last button press during a single sequence. For the set of sequences of a single type (i.e., sequence 1, 2, 3, 4, 5, or 6), we estimated the learning rate by fitting a double exponential function to the MT data \cite{Schmidt2005,Rosenbaum2010} using a robust outlier correction in {\sc Matlab} (using the function {\tt fit.m} in the Curve Fitting Toolbox with option ``Robust'' and type ``Lar''):
\begin{equation}\label{expon}
	MT = D_1 e^{-t \kappa} + D_2 e^{-t \lambda}\,,
\end{equation}
where $t$ is time, $\kappa$ is the exponential dropoff parameter (which we call the ``learning parameter'') used to describe the fast rate of improvement, $\lambda$ is the exponential dropoff parameter used to describe the slow sustained rate of improvement, and $D_1$ and $D_2$ are real and positive constants. The magnitude of $\kappa$ indicates the steepness of the learning slope: individuals with larger $\kappa$ values have a steeper dropoff in MT, suggesting that they are quicker learners \cite{Yarrow2009,Dayan2011}. The decrease in MT has been used to quantify learning for several decades \cite{Snoddy1926,Crossman1959}. Several functional forms have been suggested for the fit of MT \cite{Newell1981,Heathcote2000}, and variants of an exponential are viewed as the most statistically robust choices \cite{Heathcote2000}. Additionally, the fitting approach that we used has the advantage of estimating the rate of learning independent of initial performance or performance ceiling.


\subsection{Functional MRI (fMRI) Imaging}
\addcontentsline{toc}{subsection}{Function MRI (fMRI) Imaging}


\subsubsection*{Imaging Procedures}
\addcontentsline{toc}{subsubsection}{Imaging Procedures}

We acquired signals using a 3.0 T Siemens Trio with a 12-channel phased-array head coil. For each scan epoch, we used a single-shot echo planar imaging sequence that is sensitive to BOLD contrast to acquire 37 slices per repetition time (TR of 2000 ms, 3 mm thickness, 0.5 mm gap) with an echo time (TE) of 30 ms, a flip angle of 90 degrees, a field of view (FOV) of 192 mm, and a $64 \times 64$ acquisition matrix. Before the collection of the first functional epoch, we acquired a high-resolution T1-weighted sagittal sequence image of the whole brain (TR of 15.0 ms, TE of 4.2 ms, flip angle of 9 degrees, 3D acquisition, FOV of 256 mm, slice thickness of 0.89 mm, and $256 \times 256$ acquisition matrix).


\subsubsection*{fMRI Data Preprocessing}
\addcontentsline{toc}{subsubsection}{fMRI Data Preprocessing}

We processed and analyzed functional imaging data using Statistical Parametric Mapping (SPM8, Wellcome Trust Center for Neuroimaging and University College London, UK). We first realigned raw functional data, then coregistered it to the native T1 (normalized to the MNI-152 template with a re-sliced resolution of $3 \times 3 \times 3$ mm), and finally smoothed it using an isotropic Gaussian kernel of 8 mm full-width at half-maximum. To control for potential fluctuations in signal intensity across the scanning sessions, we normalized global intensity across all functional volumes.


\subsection{Network Construction}
\addcontentsline{toc}{subsection}{Network Construction}

\subsubsection*{Partitioning the Brain into Regions of Interest}
\addcontentsline{toc}{subsubsection}{Partitioning the Brain into Regions of Interest}

Brain function is characterized by spatial specificity: different portions of the cortex emit different, task-dependent activity patterns. To study regional specificity of the functional time series and putative interactions between brain areas, it is common to apply a standardized atlas to raw fMRI data \cite{Bassett2006b,Bassett2009b,Bullmore2009}. The choice of atlas or parcellation scheme is the topic of several recent studies in structural \cite{Bassett2010c,Zalesky2010}, resting-state \cite{Wang2009}, and task-based \cite{Power2011} network architecture. The question of the most appropriate delineation of the brain into nodes of a network is an open one and is guided by the particular scientific question at hand \cite{Bullmore2011,Wig2011}.

Consistent with previous studies of task-based functional connectivity during learning \cite{Bassett2011b,Bassett2012b,Bassett2013a,Bassett2013c,Mantzaris2013}, we parcellated the brain into 112 identifiable cortical and subcortical regions using the structural Harvard-Oxford (HO) atlas installed with the FMRIB (Oxford Centre for Functional Magnetic Resonance Imaging of the Brain) Software Library (FSL; Version 4.1.1) \cite{Smith2004,Woolrich2009}. For each individual participant and each of the 112 regions, we determined the regional mean BOLD time series by separately averaging across all of the voxels in that region.

Within each HO-atlas region, we constrained voxel selection to voxels that are located within an individual participant's gray matter. To do this, we first segmented each individual participant's T1 into white and gray matter volumes using the DARTEL toolbox supplied with SPM8.  We then restricted the gray-matter voxels to those with an intensity of 0.3 or more (the maximum intensity was 1.0). Note that units are based on an arbitrary scale. We then spatially normalized the participant T1 and corresponding gray matter volume to the MNI-152 template---using the standard SPM 12-parameter affine registration from the native images to the MNI-152 template image---and resampled to 3 mm isotropic voxels. We then restricted the voxels for each HO region by using the program fslmaths \cite{Smith2004,Woolrich2009} to include only voxels that are in the individual participant's gray-matter template.


\subsubsection*{Wavelet Decomposition}
\addcontentsline{toc}{subsubsection}{Wavelet Decomposition}

Brain function is also characterized by frequency specificity. Different cognitive and physiological functions are associated with different frequency bands, and this can be investigated using wavelets. Wavelet decompositions of fMRI time series have been applied extensively in both resting-state and task-based conditions \cite{Bullmore2003,Bullmore2004}. In both cases, they provide sensitivity for the detection of small signal changes in non-stationary time series with noisy backgrounds \cite{Brammer1998}. In particular, the maximum-overlap discrete wavelet transform (MODWT) has been used extensively in connectivity investigations of fMRI \cite{Achard2006,Bassett2006a,Achard2007,Achard2008,Bassett2009,Lynall2010}.  Accordingly, we used MODWT to decompose each regional time series into wavelet scales corresponding to specific frequency bands \cite{Percival2000}.

We were interested in quantifying high-frequency components of an fMRI signal, correlations between which might be indicative of cooperative temporal dynamics of brain activity during a task. Because our sampling frequency was 2 seconds (1 TR = 2 sec), wavelet scale one provides information on the frequency band 0.125--0.25 Hz and wavelet scale two provides information on the frequency band 0.06--0.125 Hz. Previous work has indicated that functional associations between low-frequency components of the fMRI signal (0--0.15 Hz) can be attributed to task-related functional connectivity, whereas associations between high-frequency components (0.2--0.4 Hz) cannot \cite{Sun2004}. This frequency specificity of task-relevant functional connectivity is likely due at least in part to the hemodynamic response function, which might act as a noninvertible band-pass filter on underlying neural activity \cite{Sun2004}. Consistent with our previous work \cite{Bassett2011b,Bassett2013a,Bassett2013c}, we examined wavelet scale two, which is thought to be particularly sensitive to dynamic changes in task-related functional brain architecture.


\subsubsection*{Construction of Dynamic Networks}
\addcontentsline{toc}{subsubsection}{Construction of Dynamic Networks}

For each of the 112 brain regions, we extracted the wavelet coefficients of the mean time series in temporal windows given by trial blocks (of approximately 60 TRs). The leftmost temporal boundary of each window was equal to the first TR of an experimental trial block, and the rightmost boundary was equal to the last TR in the same block. We thereby extracted block-specific data sets from the EXT, MOD, and MIN sequences (with 6--10 blocks of each sequence type) for each of the 20 participants participating in the experiment and for each of the 4 scanning sessions.

For each block-specific data set, we constructed an $N\times N$ adjacency matrix ${\bf W}$ representing the complete set of pairwise functional connections present in the brain during that window in a given participant and for a given scan. Note that $N = 112$ is the number of brain regions in the full brain atlas (see the earlier section on ``Partitioning the Brain into Regions of Interest'' for further details). To quantify the weight $W_{ij}$ of functional connectivity between regions labeled $i$ and $j$, we used the magnitude squared spectral coherence as a measure of nonlinear functional association between any two wavelet coefficient time series (consistent with our previous study \cite{Bassett2011b}). In using the coherence, which has been demonstrated to be useful in the context of fMRI neuroimaging data \cite{Sun2004}, we were able to measure frequency-specific linear relationships between time series.

To examine changes in functional brain network architecture during learning, we constructed multilayer networks by considering the set of $L$ adjacency matrices constructed from consecutive blocks of a given sequence type (EXT, MOD, or MIN) in a given participant and scanning session. We combined the matrices in each set separately to form a rank-3 adjacency tensor ${\bf A}$ per sequence type, participant, and scan.  Such a tensor can be used to represent a time-dependent network \cite{Mucha2010,Bassett2011b,Bassett2013a}.


\subsection{Network Examination}
\addcontentsline{toc}{subsection}{Network Examination}

\subsubsection*{Dynamic Community Detection}
\addcontentsline{toc}{subsubsection}{Dynamic Community Detection}

Community detection \cite{Porter2009,Fortunato2010} can be used to identify putative functional modules (i.e., sets of brain regions that exhibit similar trajectories through time). One such technique is based on the optimization of the modularity quality function \cite{NG2004,Newman2006,Newman2006b}.  This allows one to identify groups that consist of nodes that have stronger connections among themselves than they do to nodes in other groups \cite{Porter2009}. Recently, the modularity quality function has been generalized so that one can consider time-dependent or multiplex networks using \emph{multilayer modularity} \cite{Mucha2010}
\begin{equation} \label{eq:Qml}
    	Q = \frac{1}{2\mu}\sum_{ijlr}\left\{\left(A_{ijl}-\gamma_l M_{ijl} \right)\delta_{lr} + \delta_{ij}\omega_{jlr}\right\} \delta(g_{il},g_{jr})\,,
\end{equation}
where the adjacency matrix of layer $l$ has components $A_{ijl}$, the element
$M_{ijl}$ gives the components of the corresponding matrix for a null model, $\gamma_l$ is the structural resolution parameter of layer $l$, the quantity $g_{il}$ gives the community (i.e., ``module'') assignment of node $i$ in layer $l$, the quantity $g_{jr}$ gives the community assignment of node $j$ in layer $r$, the parameter $\omega_{jlr}$ is the connection strength---i.e., ``interlayer coupling parameter'', which gives an element of a tensor $\mathbf{\omega}$ that constitutes a set of \emph{temporal resolution parameters} if one is using the adjacency tensor ${\bf A}$ to represent a time-dependent network---between
node $j$ in layer $r$ and node $j$ in layer $l$, the total edge weight in the network is $\mu=\frac{1}{2}\sum_{jr} \kappa_{jr}$, the strength of node $j$ in layer $l$ is $\kappa_{jl}=k_{jl}+c_{jl}$, the intra-layer strength of node $j$ in layer $l$ is $k_{jl}$, and the inter-layer strength of node $j$ in
layer $l$ is $c_{jl} = \sum_r \omega_{jlr}$. We employ the Newman-Girvan null model within each layer by using
\begin{equation}
	M_{ijl}= \frac{k_{il} k_{jl}}{2 m_{l}}\,,
\end{equation}
where $m_{l}=\frac{1}{2} \sum_{ij} A_{ijl}$ is the total edge weight in layer $l$. We let $\omega_{jlr} \equiv \omega = \mbox{constant}$ for neighboring layers (i.e., when $| l - r | = 1$) and $\omega_{jlr} = 0$ otherwise.  We also let $\gamma_l = \gamma = \mbox{constant}$.  For simplicity and in line with previous work \cite{Bassett2011b,Bassett2013a}, here we set $\omega = 1$ and $\gamma = 1$.

\newpage
\section*{Supplementary Results}
\addcontentsline{toc}{section}{Supplementary Results}

\subsection{Description of Statistical Null Model for the Inference of Module Allegiance Values}
\addcontentsline{toc}{subsection}{Description of Statistical Null Model for the Inference of Module Allegiance Values}

In the main manuscript, we describe a method for constructing the module allegiance matrix $\mathbf{P}$, whose elements $P_{ij}$ represent the probability that area $i$ and $j$ are in the same community over a set of partitions (e.g., subjects, trial blocks, scans, etc.). This module allegiance matrix is a normalized version of the nodal association matrix described in \cite{Bassett2012b}: $\mathbf{T}$ whose elements $T_{ij}$ indicate the number of times that nodes $i$ and $j$ are in the same community. To determine which elements of $\mathbf{T}$ have probability values higher than expected, we construct a null-model association matrix $\mathbf{T^{r}}$ based on random permutations of the original set of partitions.

\begin{figure}[h]
\centerline{\includegraphics[width=1\textwidth]{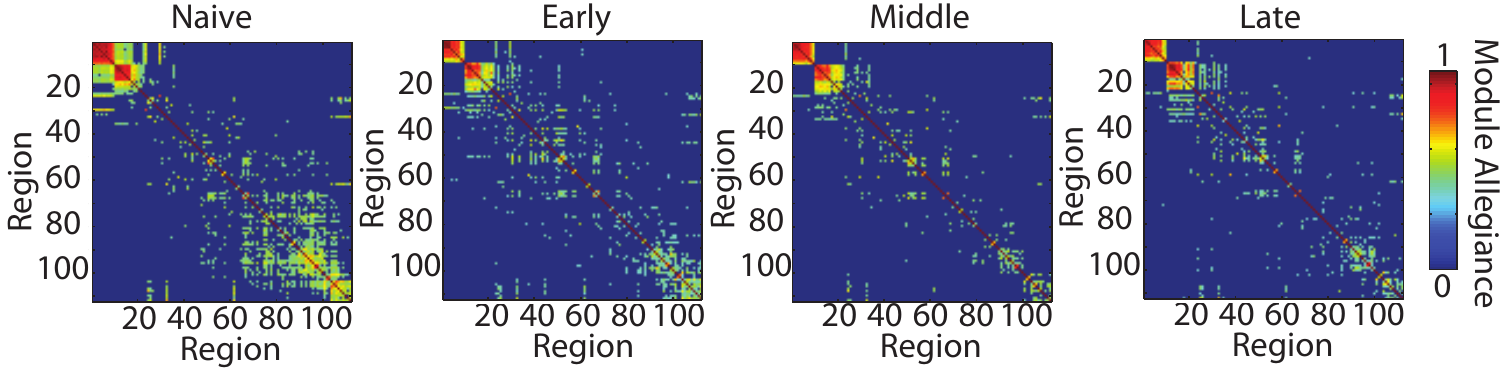}}
\caption[Dynamic Brain Architecture Associated with Task Practice Confirmed After Statistical Correction of Module Allegiance Values]{\textbf{Dynamic Brain Architecture Associated with Task Practice Confirmed After Statistical Correction of Module Allegiance Values} Here, the unthresholded module allegiance matrices given in Fig.~3 in the main manuscript are presented after the application of a statistical threshold on the module allegiance values (see Supplementary Text). We observe that the four features described in the main text -- (i) motor recruitment, (ii) visual recruitment, (iii) motor-visual integration, and (iv) non-motor, non-visual recruitment -- are preserved following this statistical correction. }\label{fig:thresh}
\end{figure}

To be concrete, let us examine the module allegiance matrices constructed for each scanning session (visualized in Fig.~3 in the main manuscript). Here, each nodal association matrix $\mathbf{T}$ is computed over the set of partitions composed of subjects and trial blocks in a single scanning session (naive, early, middle, and late). To construct an associated null model for this structure, we perform the following steps as described in \cite{Bassett2012b}. First, for each of the $C$ partitions, we reassign nodes uniformly at random to the $n$ communities that are present in the selected partition. For every pair of nodes $i$ and $j$, we let $T_{ij}^{r}$ be the number of times these two nodes have been assigned to the same community in this permuted situation. The values $T_{ij}^{r}$ then form a distribution for the expected number of times two nodes are assigned to the same partition. We observe that two nodes can be assigned to the same community a number of times out of the $C$ partitions purely by chance: the average being 38.89 times out of every 100 partitions, or 38.89\%. To be conservative, we remove such `noise' from the original nodal association matrix $\mathbf{T}$ by setting any element $T_{ij}$ whose value is less than the maximum entry of the random association matrix to $0$. This yields the thresholded matrix $\mathbf{T'}$, which retains statistically significant relationships between nodes. We then transform $\mathbf{T'}$ to a probability matrix $\mathbf{P'}$, whose elements $P_{ij}$ are either 0 (if the associated element in $\mathbf{T'}$ is 0) or the probability that area $i$ and $j$ are in the same community over the set of partitions (if the associated element in $\mathbf{T'}$ is nonzero). In Fig.~\ref{fig:thresh}, we show the thresholded $\mathbf{P'}$ matrices corresponding to the unthresholded $\mathbf{P}$ matrices visualized in Fig.~3 in the main manuscript. We observe that the four features described in the main text -- (i) motor recruitment, (ii) visual recruitment, (iii) motor-visual integration, and (iv) non-motor, non-visual recruitment -- are preserved following this statistical correction.

\subsection{Robustness of Results to Statistical Thresholding of the Module Allegiance Matrices}
\addcontentsline{toc}{subsection}{Robustness of Results to Statistical Thresholding of the Module Allegiance Matrices}

\begin{figure} [h]
\centerline{\includegraphics[width=.65\textwidth]{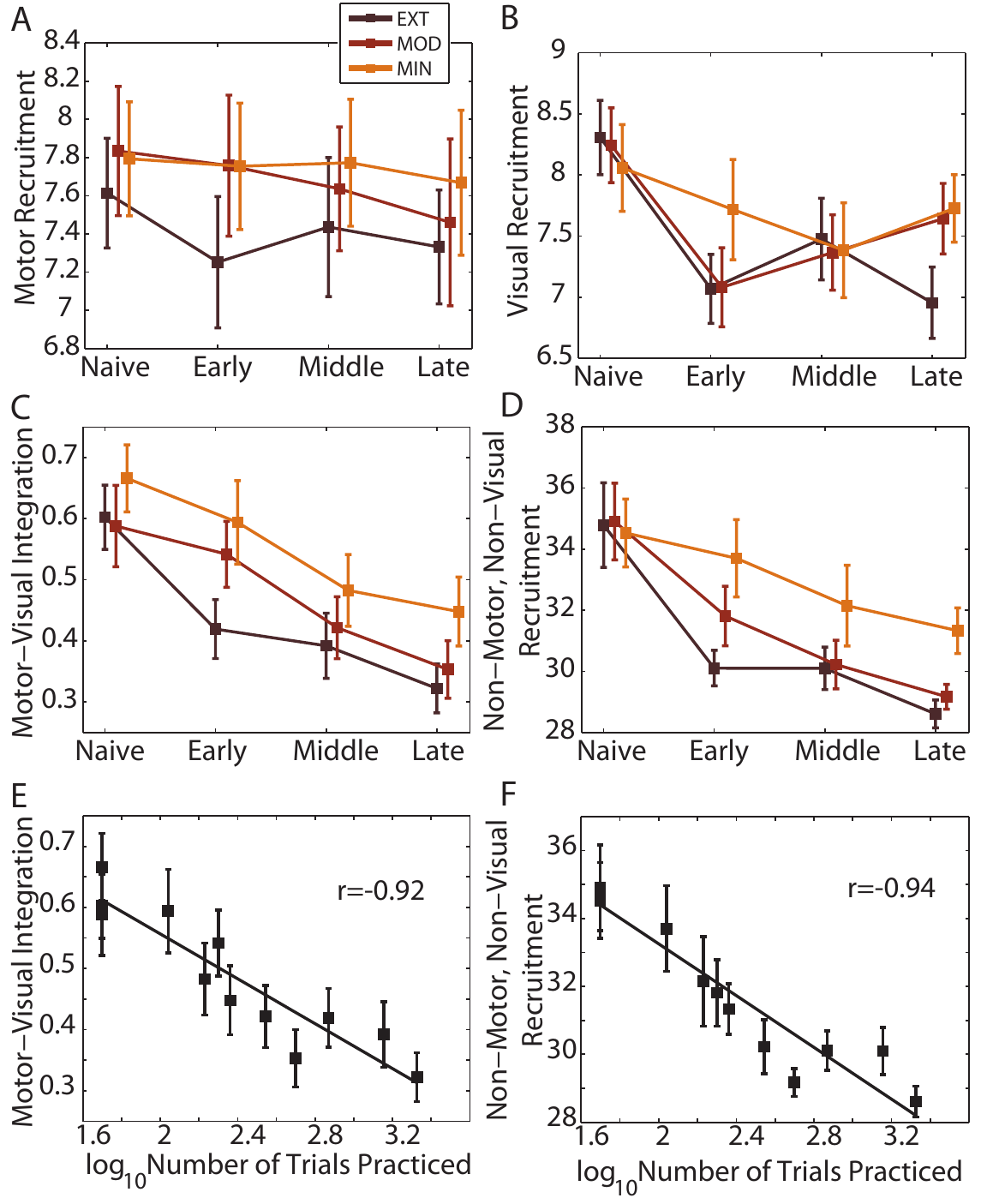}}
\caption[Recruitment and Integration Modulated by Training]{\textbf{Recruitment and Integration Modulated by Training} \emph{(A)} Motor and \emph{(B)} visual recruitment is unaffected by training intensity (extensively (maroon), moderately (red), and minimally (orange) trained sequences) and duration (naive, early, middle, and late). \emph{(C)} Integration between motor and visual modules and \emph{(D)} recruitment between non-motor and non-visual cortices decreases with training intensity and duration. \emph{(E,F)} The observations in panels \emph{(C,D)} that recruitment and integration depend on training intensity and duration can be parsimoniously described by a single latent variable: the number of trials practiced (i.e., ``depth''). Solid lines indicate best linear fit, and $r$ values indicate Pearson correlation coefficients. Error bars indicate standard deviation of the mean over participants. }\label{fig:int}
\end{figure}

\begin{figure} [h]
\centerline{\includegraphics[width=.65\textwidth]{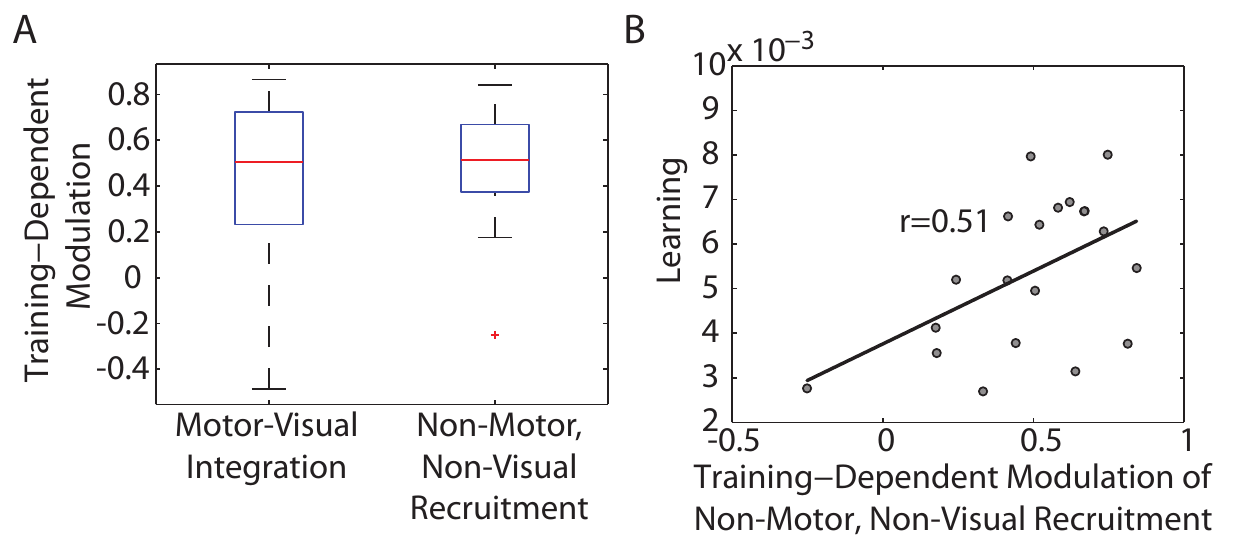}}
\caption[Individual Differences in Brain Network Architecture Map to Task Performance and Task Learning]{\textbf{Individual Differences in Brain Network Architecture Map to Task Performance and Task Learning.} \emph{(A)} Boxplots of task-dependent modulation (correlation between network diagnostic and number of trials practiced) for both motor-visual integration (left) and non-motor, non-visual recruitment (right). \emph{(B)} Scatter plot of learning and training-dependent modulation of non-motor, non-visual recruitment (Pearson's $r=0.51$).  }\label{fig:beh}
\end{figure}

In the main manuscript, we use the thresholded module allegiance matrices to compute the recruitment and integration in the motor and visual modules and the non-motor non-visual set, and to determine how these diagnostics are modulated by training. Here we demonstrate that these results are robust to the choice to statistically threshold the module allegiance matrices. Figure~\ref{fig:int} in this supplement (obtained using unthresholded module allegiance matrices)) demonstrates comparable results to those illustrated in Figure 4 of the main manuscript (obtained using statistically thresholded module allegiance matrices). Similarly, Figure~\ref{fig:beh} in this supplement (obtained using unthresholded module allegiance matrices)) demonstrates comparable results to those illustrated in Figure 5 of the main manuscript (obtained using statistically thresholded module allegiance matrices). These findings confirm that the results reported in the main manuscript are robust to the choice to statistically threshold the module allegiance matrices.

\subsection{Training-Dependent Modulation of Intra-Module Integration for Motor and Visual Systems}
\addcontentsline{toc}{subsection}{Training-Dependent Modulation of Intra-Module Integration for Motor and Visual Systems}

Here we ask whether regions in the motor or visual systems integrate further or disengage from their respective modules with increasing task practice. Following the approach we used for the modules themselves, we quantify for each subject the intra-module integration of a region to its own module as the sum of all functional connections between that region and other regions in its module. For example, the intra-module integration of the left SMA with the rest of the motor module is given by $\sum_{j (\neq i) \in \mathrm{Motor}} P_{ij}$ for $i=$left SMA and $j$ indexing the other 11 regions in the motor module. To determine if this intra-module integration of a brain region increased or decreased with task practice, we calculated the training-dependent modulation of that region: $-1 \times$ the Pearson correlation coefficient between the group-averaged intra-module integration and the number of trials practiced.

In the motor module, 7 of the 12 regions showed significant disengagement from the motor module with training, as evidenced by associated $p$-values that passed Bonferroni correction for 12 multiple comparisons: left SMA ($r=0.79$, $p=0.0023$), left precentral ($r=0.79$, $p=0.0024$), left postcentral ($r=0.80$, $p=0.0019$), left superior parietal ($r=0.77$, $p=0.0036$), right precentral ($r=0.79$, $p=0.0023$), right SMA ($r=0.80$, $p=0.0020$), left parietal operculum ($r=0.78$, $p=0.0030$). No regions showed enhanced integration with training. The disengagement of these regions is consistent with a prior study examining the functional connectivity between a handful of ROIs that reports a greater degree of connectivity early in learning than later in learning \cite{Sun2007}.

In the visual module, no regions showed significant enhanced integration or disengagement with training.

\subsection{Description of Statistical Null Model for the Skewness of the Predictive Network}
\addcontentsline{toc}{subsection}{Description of Statistical Null Model for the Skewness of the Predictive Network}

The asymmetry of strength distributions in the predictive network indicates that some brain areas participate in many functional interactions that predict learning. Based on the field of mathematics known as \emph{graph theory}, we can postulate that this level of skewness is unexpected in a purely random or noisy system. To test this hypothesis, we create an ensemble of 100,000 Erd\H{o}s-R\'{e}nyi graphs with the same number of nodes (90) and edges (180) as the predictive network. We then distribute the strength values of the edges of the predictive network uniformly at random over the edges of the Erd\H{o}s-R\'{e}nyi graph, and calculate the skewness of the strength distribution over network nodes. Using these values as a null distribution, we conclude that the observed value of skewness in the true predictive network ($s=1.89$) is significantly greater than that expected under the null hypothesis of random structure: $p=0.00009$.

\section*{Methodological Considerations}
\addcontentsline{toc}{section}{Methodological Considerations}

\subsection{Effect of Block Design}
\addcontentsline{toc}{subsection}{Effect of Block Design}

An important methodological factor is the underlying experimental block design and its effect on the coherence structure between brain regions in a single time window (i.e., in a single layer in the multilayer formalism) \cite{Bassett2013a}. Two brain regions, such as motor cortex (M1) and supplementary motor area (SMA), might be active during the trial but quiet during the inter-trial interval (ITI). This would lead to a characteristic on-off activity pattern that is highly correlated with all other regions that also turn on with the task and off during the ITI. The frequency of this task-related activity (one on-off cycle per trial, where each trial is of length 4--6 TRs) is included in our frequency band of interest (wavelet scale two, whose frequency range is 0.06--0.12 Hz), and it therefore likely plays a role in the observed correlation patterns between brain regions in a single time window.

Note, however, that our investigations of dynamic network architecture -- namely, our computations of module allegiance over the different time scales of learning -- probe functional connectivity dynamics at much larger time scales, and the associated frequencies are an order of magnitude smaller. They lie in the range 0.0083--0.012 Hz, as there is one time window every 40--60 TRs. At these longer time scales, we can probe the effects of both early learning and extended learning independently of block-design effects.

\subsection{Effect of Region Size}
\addcontentsline{toc}{subsection}{Effect of Region Size}

Recent studies have noted that brain-region size can affect estimates of hard-wired connectivity strength used in constructing structural connectomes \cite{Hagmann2008,Bassett2010c}. In prior work we demonstrated that region size does not have an appreciable effect on dynamic community structure in brain networks extracted during motor learning \cite{Bassett2013a}. However, it is nevertheless relevant to consider whether or not region size could be a driving effect of the observed organization in the module allegiance matrices.

To address this possibility, we asked whether the motor and visual modules are composed of unusually large or unusually small brain areas. We estimated the size of brain
areas in terms of voxels (averaged over participants). We then calculated the average size of all regions in the motor module (\emph{average size of motor region}), the average size of all regions in the visual module (\emph{average size of visual region}), and the average size of all regions in the non-motor, non-visual set (\emph{average size of non-motor non-visual region}). We then constructed a null distribution for each of these statistics by randomly reassigning region labels to the 3 groups (motor, visual, and non-motor non-visual). None of these groups contains significantly larger/smaller areas than expected under the null hypothesis: for the motor module $p=0.93$/$p=0.07$, visual module $p=0.30$/$p=0.70$, and non-motor non-visual set $p=0.21$/$p=0.79$. This suggests that region size is not driving the observed organization of the module allegiance matrix.

\subsection{Module Allegiance vs. Functional Connectivity}
\addcontentsline{toc}{subsection}{Module Allegiance vs. Functional Connectivity}

In the introduction of any new method, it is important to ask whether similar results could have been uncovered using a simpler approach. In this section, we demonstrate that module allegiance matrices provide a level of sensitivity to learning-related changes in brain network architecture that is not observed in functional connectivity matrices alone. We subdivide this section into discussions of (i) summary and dynamic architecture of learning, and (ii) recruitment and integration modulated by training.

\begin{figure}[h]
\centerline{\includegraphics[width=0.80\textwidth]{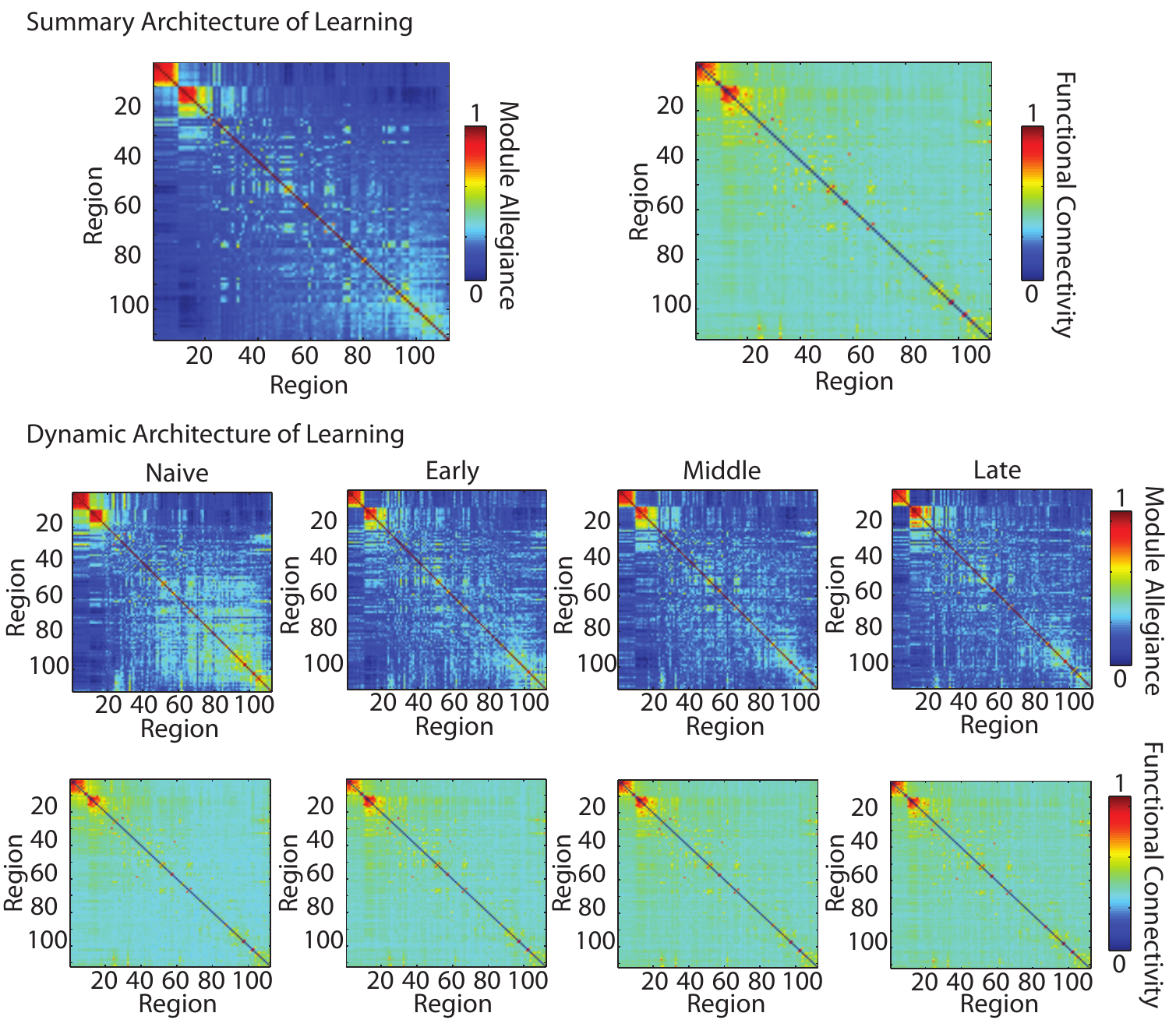}}
\caption[Module Allegiance vs. Functional Connectivity Matrices]{\textbf{Module Allegiance vs. Functional Connectivity Matrices} representing \emph{(Top)} the summary architecture of learning and \emph{(Bottom)} the dynamic architecture of learning. For comparability of visualizations, the functional connectivity matrices have been normalized by dividing by the mean.} \label{fig:fc1}
\end{figure}

\textbf{Summary and Dynamic Architecture of Learning} In Fig.~2 in the main manuscript, we show the module allegiance matrix $\mathbf{P}$, whose elements $P_{ij}$ give the probability that area $i$ and $j$ are in the same community over all subjects, scanning sessions, sequence types, and trial blocks. In the top panel of Fig.~\ref{fig:fc1}, we show the average functional connectivity matrix $\mathbf{\overline{W}}$ for comparison. The elements $W_{ij}$ give the wavelet coherence values averaged over all subjects, scanning sessions, sequence types, and trial blocks. We observe that motor and visual areas are less delineated in the average functional connectivity matrix than in the module allegiance matrix.

In Fig.~3 in the main manuscript, we show the module allegiances matrices for the naive, early, middle, and late learning sessions separately. For the naive module allegiance matrix $\mathbf{P_{\mathrm{naive}}}$, the elements $P_{ij}$ give the probability that area $i$ and $j$ are in the same community over all subjects, sequence types, and trial blocks in the naive scanning session. Module allegiance matrices for the other 3 scanning sessions are constructed similarly. In the bottom panel of Fig.~\ref{fig:fc1}, we show the average functional connectivity matrices for comparison. For the naive functional connectivity matrix $\mathbf{\overline{W}_{\mathrm{naive}}}$, the elements $W_{ij}$ give the wavelet coherence averaged over all subjects, sequence types, and trial blocks in the naive scanning session. Average functional connectivity matrices for the other 3 scanning sessions are constructed similarly. We observe that (i) motor and visual areas are less delineated, (ii) changes in motor-visual integration are less clear, and (iii) the coherent involvement of the non-motor, non-visual areas is less pronounced in the average functional connectivity matrices than in the module allegiance matrices.

\textbf{Recruitment and Integration Modulated by Training} The definitions of module recruitment and integration provided in the main manuscript utilize the module allegiance matrix $\mathbf{P}$, whose elements $P_{ij}$ give the probability that area $i$ and $j$ are in the same community. As stated in the main manuscript, we let $\mathcal{C}=\{C_1, \cdots, C_k\}$ be the partition of brain regions into groups. Then
\begin{equation}
I_{k_1, k_2} = \frac{\sum_{i\in C_{k_1}, j\in C_{k_2}}P_{ij}}{|C_{k_1}||C_{k_2}|}
\label{eq1}
\end{equation}
is the interaction strength between group $C_{k_1}$ and group $C_{k_2}$, where $|C_k|$ is the number of nodes in group $C_k$. We use this quantity to determine module recruitment and integration, depending on whether the two groups are identical ($k_1 = k_2$) or different ($k_1\neq k_2$).

It is also possible to compute alternative estimates of recruitment and integration based on the functional connectivity matrix $\mathbf{W}$, whose elements $W_{ij}$ give the wavelet coherence between the BOLD time series of area $i$ and that of area $j$. It is very important to ask whether such a substitution provides equal or better sensitivity to learning-related changes in brain network architecture.

\begin{figure}[h]
\centerline{\includegraphics[width=0.80\textwidth]{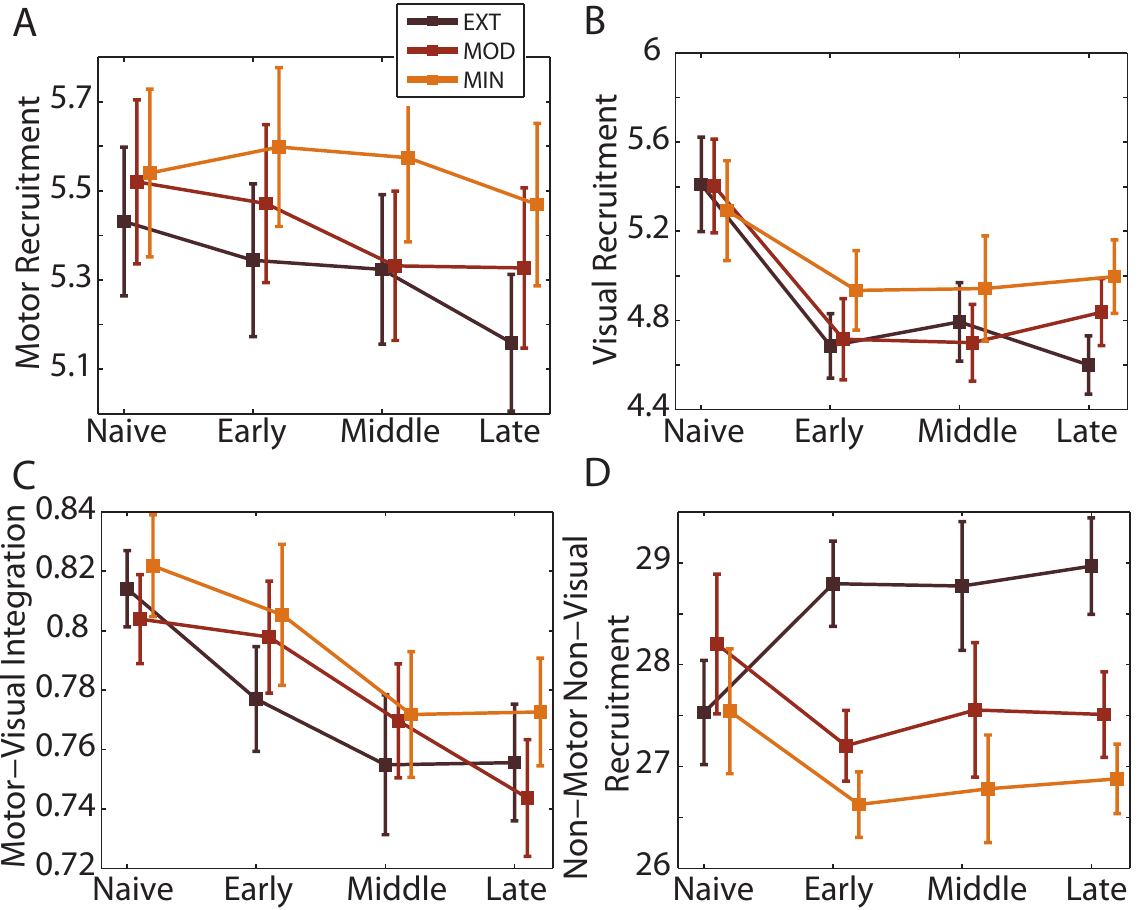}}
\caption[Alternative Estimates of Recruitment and Integration Obtained from Functional Connectivity Matrices]{\textbf{Alternative Estimates of Recruitment and Integration Obtained from Functional Connectivity Matrices} \emph{(A)} Motor, \emph{(B)} visual, and \emph{(C)} non-motor, non-visual recruitment and \emph{(D)} motor-visual integration as a function of training intensity (extensively (maroon), moderately (red), and minimally (orange) trained sequences) and duration (naive, early, middle, and late). Error bars indicate standard deviation of the mean over participants.}\label{fig:fc}
\end{figure}

To address this question, we first note that the probability matrices $\mathbf{P}$ constructed to capture the dynamic architecture of learning collapse information across trial blocks (see Fig.~4 in the main manuscript). For comparability, we therefore averaged functional connectivity matrices across trial blocks to create mean functional connectivity matrices $\mathbf{\overline{W}}$ separately for each subject, training intensity (extensively, moderately, and minimally trained sequences) and duration (naive, early, middle, and late). We then replace $P_{ij}$ in Equation 4 in this supplementary document with $\overline{W_{ij}}$, such that
\begin{equation}
I^{\prime}_{k_1, k_2} = \frac{\sum_{i\in C_{k_1}, j\in C_{k_2}}\overline{W_{ij}}}{|C_{k_1}||C_{k_2}|}
\label{eq2}
\end{equation}
is an alternative estimate of the interaction strength. In Fig.~\ref{fig:fc}, we show the values of the four brain network diagnostics (motor, visual, and non-motor non-visual recruitment and motor-visual integration) computed using $I^{\prime}$ (Equation 5) rather than $I$ (Equation 4). We observe that none of these diagnostics display training-dependent modulation consistent with the number of trials practiced. Instead, we tend to observe relatively independent effects of either training intensity (see Fig.~\ref{fig:fc}A \& D) or training duration (see Fig.~\ref{fig:fc}B), or neither (see Fig.~\ref{fig:fc}C).

\clearpage
\newpage

\small
\begin{table}
\begin{center}
\begin{tabular}{| l | l |}
\hline
Frontal pole &	Cingulate gyrus, anterior \\
Insular cortex&	Cingulate gyrus, posterior \\
Superior frontal gyrus & Precuneus cortex \\
Middle frontal gyrus & Cuneus cortex \\
Inferior frontal gyrus, pars triangularis & Orbital frontal cortex \\
Inferior frontal gyrus, pars opercularis &Parahippocampal gyrus, anterior \\
Precentral gyrus &Parahippocampal gyrus, posterior \\
Temporal pole & Lingual gyrus \\
Superior temporal gyrus, anterior &Temporal fusiform cortex, anterior \\
Superior temporal gyrus, posterior& Temporal fusiform cortex, posterior \\
Middle temporal gyrus, anterior &Temporal occipital fusiform cortex \\
Middle temporal gyrus, posterior &Occipital fusiform gyrus \\
Middle temporal gyrus, temporooccipital& Frontal operculum cortex \\
Inferior temporal gyrus, anterior & Central opercular cortex \\
Inferior temporal gyrus, posterior & Parietal operculum cortex \\
Inferior temporal gyrus, temporooccipital 	&Planum polare \\
Postcentral gyrus	&Heschl's gyrus \\
Superior parietal lobule	&Planum temporale \\
Supramarginal gyrus, anterior&	Supercalcarine cortex \\
Supramarginal gyrus, posterior	&Occipital pole \\
Angular gyrus	&Caudate \\
Lateral occipital cortex, superior	&Putamen \\
Lateral occipital cortex, inferior	&Globus pallidus \\
Intracalcarine cortex	&Thalamus \\
Frontal medial cortex	&Nucleus Accumbens \\
Supplemental motor area&	Parahippocampal gyrus (superior to ROIs 34,35) \\
Subcallosal cortex	&Hippocampus \\
Paracingulate gyrus	&Brainstem \\
\hline
\end{tabular}
\end{center}
\caption[Brain regions present in the Harvard-Oxford Cortical and Subcortical Parcellation Scheme]{\footnotesize \textbf{Brain regions present in the Harvard-Oxford Cortical and Subcortical Parcellation Scheme provided by FSL \cite{Smith2004,Woolrich2009}.}  \label{Table1}}
\end{table}
\normalsize

\clearpage
\newpage
\bibliography{bibfile}